\documentclass{emulateapj}

\newcommand{\vdag}{^\dagger}

\usepackage{pifont}

\shorttitle{NIR spectra of background stars and low mass YSOs.}
\shortauthors{Noble et al.}

\begin{document}

\title{A survey of H$_2$O, CO$_2$ and CO ice features towards background stars and low mass YSOs using AKARI.\altaffilmark{1}}

\author{J.~A. Noble\altaffilmark{2}}
\affil{Aix-Marseille Universit\'{e}, PIIM UMR 7345, 13397, Marseille, France.}
\author{H.~J. Fraser\altaffilmark{2}}
\affil{Department of Physical Sciences, The Open University, Walton Hall, Milton Keynes MK7 6AA, United Kingdom.}
\email{helen.fraser@open.ac.uk}
\author{Y. Aikawa}
\affil{Department of Earth and Planetary Sciences, Kobe University, Kobe 657-8501, Japan.}
\author{K.~M. Pontoppidan}
\affil{Space Telescope Science Institute, 3700 San Martin Drive, Baltimore, MD 21218, U.S.A.}
\and
\author{I. Sakon}
\affil{Department of Astronomy, Graduate School of Science, University of Tokyo, 7-3-1 Hongo, Bunkyo-ku, Tokyo 113-0003, Japan.}

\altaffiltext{1}{Based on observations with AKARI, a JAXA project with the participation of ESA.}
\altaffiltext{2}{Previous address: Department of Physics, Scottish Universities Physics Alliance, University of Strathclyde, Glasgow, G4 ONG, Scotland.}

\begin{abstract}

We present near infrared spectroscopic observations of 19 molecular clouds made using the AKARI satellite, and the data reduction pipeline written to analyse those observations. The 2.5~--~5~$\mu$m spectra of 30 objects -- 22 field stars behind quiescent molecular clouds and eight low mass YSOs in cores -- were successfully extracted using the pipeline. Those spectra are further analysed to calculate the column densities of key solid phase molecular species, including H$_2$O, CO$_2$, CO, and OCN$^-$. The profile of the H$_2$O ice band is seen to vary across the objects observed and we suggest that the extended red wing may be an evolutionary indicator of both dust and ice mantle properties. The observation of 22 spectra with fluxes as low as $<$~5~mJy towards background stars, including 15 where the column densities of H$_2$O, CO and CO$_2$ were calculated, provides valuable data that could help to benchmark the initial conditions in star-forming regions prior to the onset of star formation.

\end{abstract}


\section{Introduction}
The three most abundant solid phase molecular species in the interstellar medium (ISM) are H$_2$O, CO$_2$, and CO. The strongest absorptions of all three species are seen in the 2.5~--~5~$\mu$m region of the spectrum, associated with the stretching mode vibrations of each molecule. AKARI was the first space telescope since ISO with which it was possible to make spectroscopic observations of the full near infrared (NIR) range. Compared with \emph{Spitzer} or ground-based observations, AKARI spectroscopy offers two distinct advantages: simultaneous observation of the stretching mode absorption bands of H$_2$O, CO$_2$, and CO, and observation of the blue wing of the H$_2$O stretching mode absorption. Two further benefits of AKARI are slitless spectroscopy, with spectra simultaneously captured for all objects in the field of view, and exceptional sensitivity, allowing observation of ice spectra on lines of sight towards objects with fluxes approaching 1~mJy. 

We present a study of the 2.5 -- 5~$\mu$m ice absorption spectra towards background stars and young stellar objects (YSOs) with AKARI. To date, very few spectroscopic ice studies with AKARI have been published \citep[see e.g.][]{shi08,shi10,bur10,yam11,Aikawa12,Sorahana12} reflecting the difficulty of reducing AKARI spectral data. In all previous observations, ice spectra were derived from a single, relatively bright, object in the field of view. To study background star ice spectra, we have extracted data from multiple faint objects in a single observation, requiring the development of a unique data reduction pipeline. 

The ultimate aim of our AKARI observing programme is to map the spatial distribution of the key ice species H$_2$O, CO$_2$, and CO. A prerequisite, however, is a robust data reduction and analysis method, to ensure the reliable determination of molecular ice column densities over hundreds of objects concurrently. The first aim of this article is to detail this analysis process. These data enhance the sample of ice spectra towards background stars and low mass YSOs; our second aim is to test how the chemical processes governing ice formation and evolution are related.

The chemistry of CO, CO$_2$, and H$_2$O in molecular clouds is intrinsically linked. Solid H$_2$O is the dominant interstellar ice species, seemingly ubiquitous in the dense, molecular ISM \citep{Leger79,Whittet83,Boogert11}. This H$_2$O forms on the surface of silicate dust grains \citep{Tielens82,Boogert04,Miyauchi08,Ioppolo08,Duli10}, most likely through a combination of reactions involving H atoms, O atoms, and OH radicals.

Gas phase CO is highly abundant in molecular clouds and is commonly used to trace astrophysical conditions \citep[e.g.][]{vankemp99}. However, unlike H$_2$O, CO accretes onto icy grain surfaces at a rate proportional to the gas density \citep[e.g.][]{aik01,pon03}, resulting in a CO-rich ice layer. Subsequently, as a function of time, temperature, or other energetic inputs to the ice mantle, the H$_2$O and CO layers can mix, producing a second CO ice environment, as illustrated by previous experimental \citep{collings03} and observational \citep{pon03} work. Both of these CO ice environments have been detected in interstellar observations \citep{tie91,Whittet91,Chiar95,pon03}. 

CO$_2$ is believed to form in the solid phase, based upon its low gas phase abundances \citep{vanD96}. CO$_2$ ice has been observed towards numerous interstellar environments, including low mass YSOs \citep{num01,pon08} and background stars \citep{kne05}. Laboratory experiments prove that CO$_2$ formation is effective via several energetic \citep{ger96,pal98,Jamieson06,men04} and purely thermal \citep{gou08,Oba10,iop11a,nob11,zins11} mechanisms. Recent results imply that the reaction CO + OH occurs in competition with H$_2$O formation (via H + OH), and that some CO molecules freeze-out from the gas phase long before CO ice is formed. Such postulations can only be tested through concurrent observations of CO, CO$_2$, and H$_2$O ices.

Any sample of ice spectra observed to date has been, necessarily, dictated by both the spectral range and sensitivity of IR telescopes. Low mass YSOs have therefore been rather more extensively studied than background stars, beginning with limited ISO observations \citep{Guertler96,deG96,ger99,Alexander03}. \emph{Spitzer}, more sensitive than ISO, observed many more low mass YSOs \citep[e.g.][]{gib04,Zasowski09,Boogert08}. The \emph{Spitzer} Legacy program ``From Molecular Cores to Planet-Forming Disks'' (c2d) extensively detailed the abundances of CO$_2$ \citep{pon08}, H$_2$O \citep{Boogert08}, CH$_4$ \citep{Oberg08}, and NH$_3$ \citep{Bott10} ices towards a sample of 41 low mass YSOs, providing the largest dataset of column density values for these key interstellar molecules \citep{Oberg11}. An important conclusion of this survey is that the abundances of the major ice species are remarkably similar in different star forming regions, although, for reasons not yet fully determined, the ice composition of minor and volatile species varies extensively. This suggests that ices form in the quiescent ISM, a region probed by ice spectra towards background stars.

\begin{deluxetable}{lccc}
 \tabletypesize{\footnotesize}
\tablewidth{0pt}
\tablecaption{Molecular cores observed \label{tbl-1}}
\tablehead{         & Coordinates                                  & Observation             & Date\\
Core                    & R. A., Dec. /deg\tablenotemark{a} & ID\tablenotemark{b} & dd/mm/yy}
\startdata
L~1551                & 67.8708, +18.1356       & 3010019\_001              & 28/02/07\\
B~35A                & 86.126, +9.1461             & 4120022\_001$\vdag$ & 16/03/07\\
DC~269.4+03.0 & 140.5938, -45.7896       & 4120043\_001               & 09/12/06\\
                         &                                         & 4121043\_001                & 08/06/07\\
DC~274.2-00.4 & 142.2042, -51.6153     & 4120007\_001$\vdag$    & 17/12/06\\
DC~275.9+01.9 & 146.6904, -51.1000    & 4120009\_001$\vdag$     & 20/12/06\\
DC~291.0-03.5 & 164.9663, -63.7227    & 4120023\_001$\vdag$     & 21/01/07\\
BHR~59             & 166.7875, -62.0994    & 4120002\_001$\vdag$    & 19/01/07\\
DC~300.2-03.5 & 186.0904, -66.2024    & 4121045\_001$\vdag$    & 07/08/07\\
Mu~8                 & 187.7146, -71.0227    & 4120018\_001$\vdag$    & 12/02/07\\
DC~300.7-01.0 & 187.8771, -63.7233    & 4120024\_001                 & 02/02/07\\
                         &                                        & 4121024\_001                  & 05/08/07\\
BHR~78             & 189.0583, -63.1920    & 4121041\_001$\vdag$     & 05/08/07\\
B~228               & 235.6744, -34.1621    & 3011017\_001                   & 24/08/07\\
BHR~144          & 249.3750, -35.2333    & 3010017\_001                   & 03/03/07\\
L~1782-2        & 250.6333, -19.7272    & 3010010\_001                   & 03/03/07\\
LM~226            & 254.3333, -16.1461    & 3010012\_001                   & 06/03/07\\
L~1082A            & 313.3792, +60.2464    & 3010028\_001                  & 22/12/06\\
L~1228             & 314.3956, +77.6282    & 3011006\_001                 & 22/08/07\\
L~1165             & 331.6683, +59.0999    & 4121035\_001$\vdag$   & 07/07/07\\
L~1221            & 337.1083, +69.0417    & 3010003\_001                & 26/01/07\\
\enddata 
\tablenotetext{a}{Right Ascension and Declination (J2000) at the centre of the Np aperture.}
\tablenotetext{b}{Observations where a second pointing with ID XXXXXXX\_002 was made immediately after the first are represented by $\dag$. All other observations are explicitly tabulated.}
\end{deluxetable}

\begin{figure*}[htb]
  \plotone{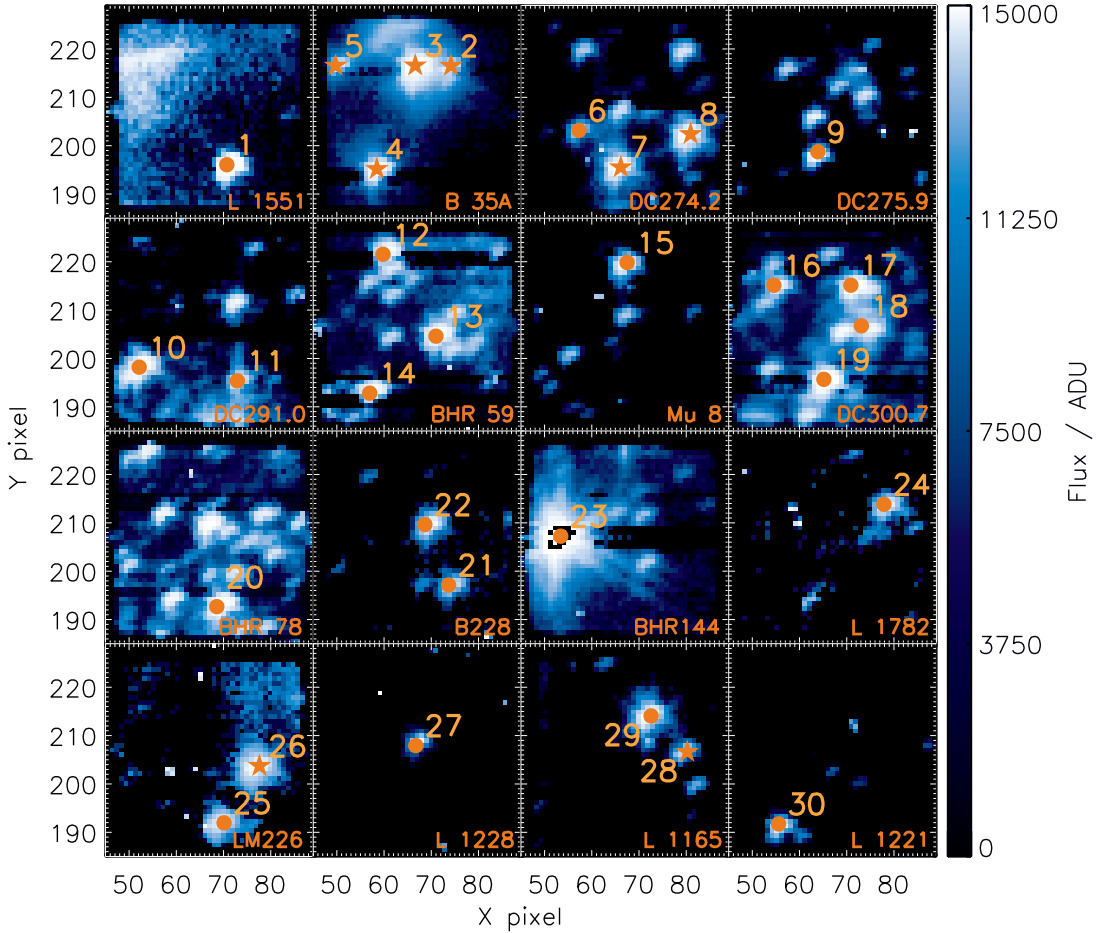}
  \caption{Imaging frames (at 3.2~$\mu$m) for the 1'~$\times$~1' grism field of view for the 16 cores that included extractable objects. X pixel is aligned with right ascension, Y pixel with declination, and the colourbar indicates the flux in each pixel. The 30 objects for which ice spectra were extracted are numbered in order of ascending right ascension; this sample includes 22 background stars ($\bullet$) and eight YSOs ($\star$). Details of each source object, including J, H, K$_S$ photometric values and object coordinates are given in Table~\ref{tbl-3}. For full details on the selection of cores and objects see \S~\ref{sec-target-selection} and \S~\ref{sec-classification}, respectively.\label{fig-fovs}}
\end{figure*}

Limited background stars observations were first made in the 2.4~--~5~$\mu$m wavelength range by ISO; more recently, \emph{Spitzer} has observed many more background stars, but not in the 2.5~--~5~$\mu$m range covering the stretching mode absorptions of H$_2$O, CO$_2$, and CO. Observing ices in this region is vital in benchmarking the inventory of ice species present at the earliest stages of star formation. Two background stars were observed by ISO: Elias 13 was observed in the range 4.20~--~4.34~$\mu$m (focusing on the CO$_2$ stretching mode at 4.25~$\mu$m \citep{num01}) whilst Elias 16 was observed in the ranges 2.4~--~5~$\mu$m \citep{Whittet98,gib04} and 14~--19~$\mu$m, with additional detailed studies of the CO$_2$ stretching mode \citep{Whittet98,ger99,num01}, $^{13}$CO$_2$ \citep{Boogert2000}, and potentially the N$_2$ stretching mode \citep{Sandford01}. With \emph{Spitzer}, mid IR observations were made of the same two background stars (Elias 13 (5~--~20~$\mu$m) and Elias 16 (5~--~14~$\mu$m)) and a third background object -- CK~2 (5~--~20~$\mu$m) \citep{kne05}. The CO$_2$ bending mode region around 15.2~$\mu$m was investigated for a further 12 background stars across Taurus, Serpens and IC~5146 \citep{ber05,Whittet07,Whittet09}. CO$_2$ was found to exist in H$_2$O-poor ices in quiescent regions for the first time \citep{ber05}, pointing to a formation mechanism involving CO ice. Successful ground-based studies have been carried out to observe the 4.67~$\mu$m CO stretch towards 12 background stars \citep{Whittet85,Whittet89,Eiroa89} and the observable part of the 3~$\mu$m water band towards 73 background stars \citep{Whittet88,Eiroa89,Murakawa00}. By combining ground-based and \emph{Spitzer} observations, \citet{Boogert11} quantified the ice abundances of H$_2$O, NH$_4^+$, CH$_3$OH, HCOOH, CH$_4$, and NH$_3$ towards 31 background stars (also including CO$_2$ abundances towards eight objects) and confirmed that little difference in the abundances of major ice species is observed between background stars and YSOs. 

Most previous background star studies have been made towards Taurus and Serpens molecular clouds, with the exception of two studies of the Cocoon Nebula (IC~5146) \citep{Whittet09,Chiar11}, one of $\rho$ Ophiuchi \citep{Tanaka90}, and the survey by \citet{Boogert11}, which observed 31 background stars towards 16 isolated dense cores. Our study adds 22 background stars across a further 15 dense molecular cores to these statistics. \citet{Boogert11} highlight that the existing studies of background stars focus on cores and cloud regions well separated from lines of sight where YSOs have been studied. To understand the subtle changes in ice chemistry as a region evolves throughout the star-formation process would require comparisons between ice spectra of these different evolutionary objects towards the same core. In three of our cores, it has been possible to compare ice column densities from background and YSO sources separated by only a few 1000 AU.

In this paper, the column densities of H$_2$O, CO$_2$, and CO are calculated from the 2.5~--~5~$\mu$m AKARI spectra of 22 background stars and eight low mass YSOs. As the only complete spectrum of a background star in this spectral region to date is that of Elias 16 taken by ISO, this represents a significant increase in data on background stars in the NIR. In addition to H$_2$O, CO$_2$, and CO, it was possible in certain cases to calculate the column densities of OCN$^-$, and CO gas-grain \citep{fraser05}. This large increase in data, including low flux background stars, offers much evidence about the initial conditions in quiescent molecular clouds before star formation begins. In particular, it offers clues to where ices form, and how the chemistry of different molecular species is linked in prestellar regions.

 \begin{figure*}[htb]
   \plotone{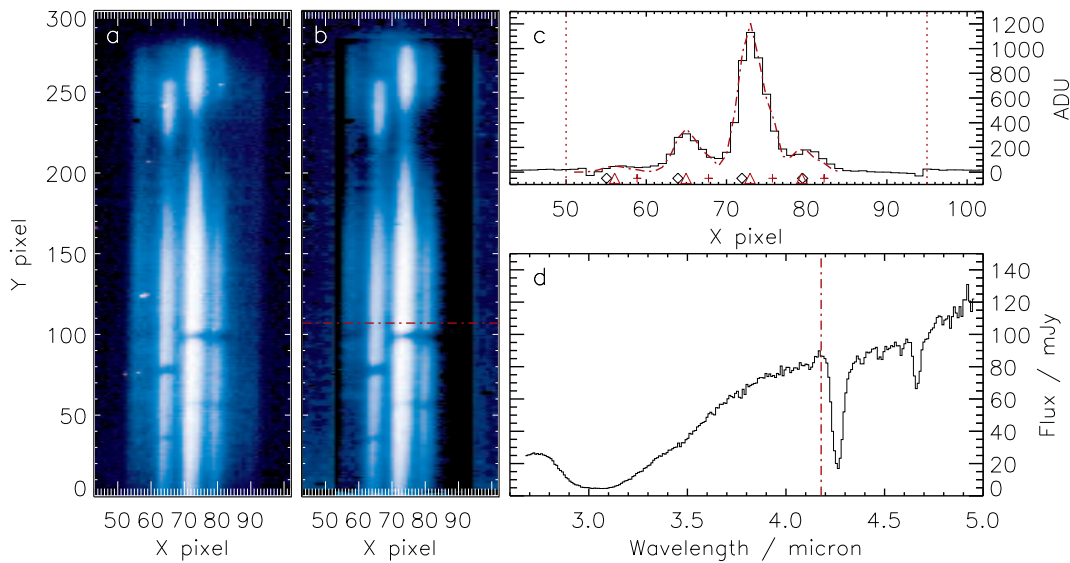}
   \caption{An illustration of our ``AKARI Reduction Facility'' pipeline, using an observation of molecular cloud B~35A. a) Single raw AKARI spectral data frame from the 1'~$\times$~1' region of the NIR detector. The light from the region of sky in the 1'~$\times$~1' field of view (see Figure~\ref{fig-fovs}) is dispersed by the grism disperser along the Y axis. The X axis remains a spatial dimension, while the Y axis is both a spatial and dispersion direction. In the Z direction (represented here by colour intensity, as per the colourbar in Figure~\ref{fig-fovs}) is the flux information, in Automated Data Units (ADU). b) Fully corrected data frame. Eight spectral frames have been dark-subtracted, corrected for hot and bad pixels, shifted and stacked to produce a final, corrected spectral frame. Compared with (a), in this sharper image the dispersed spectra and the ice absorptions are more distinct. Extraction of spectral data proceeds row by row along the dispersion direction; the dot-dashed (red) line represents the row Y=108. c) The black line shows the flux ``slice'' (in ADU) versus X pixel along row Y=108 (dot-dashed (red) line in (b)). The data are fitted with a series of Gaussian functions, as described in \S~\ref{sec-extraction}. In this example, there are four objects being fitted (whose X positions are represented by black diamonds on the X axis). The central positions of the Gaussian functions in the fit are plotted as (red) triangles (on the X axis); the overall fit is superimposed as a dot-dashed (red) line. This fit well describes the data and deconvolves the overlapping flux of the four distinct objects. This process is repeated for every row in the Y direction. d) Overall final spectrum of the brightest object in this field of view. The spectrum is reconstructed from the series of flux datapoints extracted by the row fitting illustrated in (c); the datapoint produced from the fit in (c) is represented by the dot-dashed (red) line. The final spectrum on a flux-wavelength scale is produced by division by the spectral response function.\label{fig-pipeline}}
 \end{figure*}

\section{Observations}

Observations were carried out between December 2006 and August 2007 using the spectroscopic mode of the Infrared Camera on AKARI \citep{ona07,ohy07}. We observed fields of view towards 19 dense molecular clouds. Of these, 11 were part of the AKARI European Users Open Time Observing Programme IMAPE, a selection identified from the c2d \emph{Spitzer} Legacy program \citep{eva03}; a further eight fields of view are included here from the Japan/Korea Users Open Time Observing Programme ISICE.  Details of the 19 cores are given in Table~\ref{tbl-1}. 

In the IMAPE programme, clouds were selected by identifying those containing dense prestellar cores, a statistically large number of identifiable field stars behind the cloud, and with the proviso that no previously identified YSO candidates were in the line of sight. Two pointings were made towards each cloud, affording an improvement in the spectral signal-to-noise, and allowing for better removal of hot pixel and cosmic ray effects from the data. Details of the cloud selections in the ISICE programme, for which only single pointings were made of each cloud, are given in \citet{Aikawa12}.

The observations presented here focus specifically on data obtained with the IRC04 AKARI astronomical observing template (AOT) using the NIR grism disperser (NG). The AOT produced eight spectral frames, an imaging frame and a pre- and post- dark frame per field of view per pointing. The spectra were extracted from the Np aperture, a 1'~$\times$~1' field of view with a spectral range 2.5~--~5.0~$\mu$m (R = 120 at 3.6~$\mu$m). Imaging frames for 16 of the clouds are shown in Figure~\ref{fig-fovs}. The images were taken with the N3 filter (2.7~--~3.8~$\mu$m) at a reference wavelength of 3.2~$\mu$m. As is clear from Figure~\ref{fig-fovs}, some fields of view contain multiple objects, and as the AKARI spectroscopic mode is slitless, the spectral traces of these objects often overlap. Nevertheless, 30 individual objects were identified across these fields of view (highlighted in Figure~\ref{fig-fovs}) for which spectra were successfully extracted using our data reduction pipeline. The selection and classification of these individual objects are discussed later in \S~\ref{sec-target-selection}. As far as we can ascertain from the literature, ice spectra have not previously been published for any of the objects in the lines of sight observed here. Three clouds listed in Table~\ref{tbl-1} do not feature again in this paper. The field of view in DC 300.2-03.5 was too confused to enable us to extract any spectra, even though ice features may have been present; data from L 1082A and DC 269.4+03.0 were below the lower flux limit at which our pipeline functioned effectively, as discussed in \S~\ref{sec-target-selection}.

\section{Data reduction}\label{sec-reduction}

\begin{deluxetable*}{cccccccc}
\tabletypesize{\scriptsize}
\tablewidth{0pt}
\tablecaption{Source list.\tablenotemark{a}\label{tbl-2}}
\tablehead{
                                                       &                                       &  R. A.                    & Dec.                     &                         & J                       & H                      & K$_S$ \\      
\colhead{Object\tablenotemark{b}} & \colhead{2MASS Name} & \colhead{(J2000)} & \colhead{(J2000)} & \colhead{Core} & \colhead{(mJy)} & \colhead{(mJy)} & \colhead{(mJy)}  }
\startdata
1$\vdag$ & 04312835+1807427 & 67.868103 & 18.128599 & L~1551        & 0.143   & 1.170  & 2.676    \\
2 & J054429.3+090857\tablenotemark{c} & 86.122498 & 9.1490002  & B~35A       & \nodata & \nodata & \nodata \\ 
3 & 05443000+0908573 & 86.124964 & 9.1492458  & B~35A                & 0.067   & 0.442   & 7.310   \\ 
4 & 05443085+0908260 & 86.128544 & 9.1406136  & B~35A                & 0.0701  & 0.537  & 5.420   \\ 
5$\vdag$& 05443164+0908578 & 86.131846 & 9.1494267  & B~35A                & 0.393   & 1.320   & 2.370   \\ 
6$\vdag$  & 09284840-5136379 & 142.20166 & -51.610514 & DC~274.2-00.4  & 0.086   & 0.373  & 1.880  \\ 
7  & 09285020-5136373 & 142.20929 & -51.610411 & DC~274.2-00.4  & 2.380   & 8.000  & 13.600 \\ 
8 & 09285128-5136588 & 142.21379 & -51.616373 & DC~274.2-00.4  & 3.280   & 9.080  & 13.900 \\ 
9 & 09464633-5105424 & 146.69306 & -51.095141 & DC~275.9+01.9 & 5.530   & 6.260  & 4.580  \\ 
10 & 10595211-6343000 & 164.96722 & -63.716741 & DC~291.0-03.5& 10.300  & 15.700  & 15.600 \\ 
11 & 10595548-6343210 & 164.98119 & -63.722508 & DC~291.0-03.5 & 1.150   & 2.320  & 2.370  \\ 
12  & 11070622-6206057 & 166.77597 & -62.101584 & BHR~59              & 0.705   & 4.190   & 7.760  \\    
13  & 11071041-6206013 & 166.79341 & -62.100358 & BHR~59              & 0.804   & 5.380   & 11.900 \\   
14  & 11071034-6205345 & 166.79304 & -62.092943 & BHR~59               & 0.0997 & 1.140  & 6.150  \\   
15 & 12305017-7101331 & 187.70904 & -71.025906 & Mu~8                 & 4.370   & 9.270  & 9.890  \\ 
16 & 12312792-6343221 & 187.86635 & -63.722816 & DC~300.7-01.0 & 1.370   & 5.090  & 7.230  \\ 
17$\vdag$ & 12313060-6343376 & 187.87753 & -63.727139 & DC~300.7-01.0 & 3.960   & 16.400 & 24.700 \\ 
18$\vdag$ & 12313217-6343306 & 187.88413 & -63.725144 & DC~300.7-01.0 & 0.718   & 4.370   & 7.250  \\ 
19 & 12313247-6343105 & 187.88539 & -63.719644 & DC~300.7-01.0 & 4.230   & 24.800 & 44.200 \\ 
20 & 12361240-6311509 & 189.05177 & -63.197538 & BHR~78           & 0.488   & 3.030   & 8.500  \\ 
21$\vdag$ & 15424095-3409598 & 235.67059 & -34.166599 & B~228              & 0.225   & 1.376   & 1.916  \\ 
22$\vdag$ & 15424185-3409435 & 235.67439 & -34.162102 & B~228              & 0.905   & 3.406   & 4.974  \\
23$\vdag$ & 16372876-3513588 & 249.36990 & -35.233002 & BHR~144          & 13.125  & 121.759 & 331.688\\
24$\vdag$ & 16423341-1943501 & 250.63921 & -19.730600 & L~1782-2   & 0.772   & 1.895   & 2.488  \\
25$\vdag$ & 16572088-1608241 & 254.33701 & -16.139999 & LM~226           & 0.400   & 2.004   & 3.594  \\
26$\vdag$ & 16572151-1608423& 254.33960 & -16.145100 & LM~226           & 0.625   & 3.447   & 7.833  \\
27$\vdag$ & 20573495+7737415 & 314.39560 & 77.628197  & L~1228            & 1.014   & 4.737   & 7.783 \\
28  & 22064175+5906156 & 331.67388 & 59.104326  & L~1165            & 0.137   & 0.452   & 1.490  \\ 
29  & 22064185+5906000 & 331.67432 & 59.100005  & L~1165            & 0.216   & 3.470   & 12.300 \\ 
30$\vdag$  & 22283131+6902272 & 337.13049 & 69.040901  & L~1221      & 0.785   & 3.483   & 5.057  \\
\enddata
\tablenotetext{a}{This table contains properties and extraction details of all objects with a spectrum or an upper limit. As seen in Figure~\ref{fig-fovs}, objects are numbered in order of ascending RA. Here, the positional and photometric properties of these objects are detailed.}
\tablenotetext{b}{The final spectra of all objects marked with $\dag$ were created from only one extracted spectrum.}
\tablenotetext{c}{This object does not have a 2MASS ID and is thus labelled with its \emph{Spitzer} c2d ID. Neither does it have J, H, K$_S$ band flux values.}
\end{deluxetable*}

\subsection{AKARI Reduction Facility (ARF)}

The AKARI 1'~$\times$~1' aperture was designed for point source spectroscopy of single sources and the data reduction toolkit written by the IRC team (\citet{ohy07}, hereafter referred to as ``the toolkit'') was intended to analyse such sources. 
In the majority of the observations in this work, there was more than one object in the 1'~$\times$~1' aperture; the dispersed spectra of these objects overlapped and produced confused spectra when reduced using the toolkit. A purpose written data reduction pipeline -- the AKARI Reduction Facility, hereafter referred to as ``ARF'' -- was therefore developed by the authors to analyse AKARI NIR spectroscopic data. ARF was written in IDL, incorporating procedures taken from the IDL Astronomy User's Library \citep{lan03} and the Markwardt Library \citep{mar09}. Full details of ARF are presented in \citet{NobleThesis}.

The basic function of ARF is illustrated in Figure~\ref{fig-pipeline}: the eight raw spectral data frames (Figure~\ref{fig-pipeline}a) are dark-subtracted (using the pre-dark frame) and corrected for bad and hot pixels before being shifted and stacked. As with previously published AKARI spectroscopic studies, no flat-fielding of the spectral frames is performed, as the calibration data provided in the toolkit did not include flat fields for this region of the detector. The shift between frames is calculated by cross-correlation, for the direction perpendicular to dispersion only (X). Frames are shifted by non-integer values, as appropriate, with linear interpolation of flux to maintain absolute flux values upon stacking. Background correction is performed on the stacked frame, using a region of dark sky picked manually from the spectral aperture (Figure~\ref{fig-pipeline}b). Objects in the field of view are identified according to their coordinates, taking account of the distortion in the frame. The spectrum of each object is then extracted by fitting the grism point spread function (PSF) to the dispersed flux (Figure~\ref{fig-pipeline}c). Extracted data are divided by the spectral response function (SRF) of the NIR detector to produce spectra on a flux-wavelength scale (Figure~\ref{fig-pipeline}d). In general, this is a standard reduction technique, the main exception being the extraction method, described in detail below. 

\subsection{Extraction of spectra}\label{sec-extraction}

\subsubsection{Target object specification}\label{sec-target-selection}

Objects for extraction are introduced to ARF from target tables. Pointings that were part of the IMAPE program are a subset of the c2d \emph{Spitzer} Legacy program and had target tables available from that program \citep{eva03}, including photometric data points from the Two Micron All Sky Survey (2MASS) and \emph{Spitzer} InfraRed Array Camera (IRAC). In ARF, the c2d target lists are edited to exclude objects outside the 1'~$\times$~1' aperture and those below a threshold 2MASS K$_S$ band (centred at 2.2~$\mu$m) magnitude of 1.19~mJy, the magnitude of the brightest isolated object (in DC~269.4+03.0) which could not be fitted with the PSF using ARF (discussed in \S~\ref{psf}). Target tables were prepared by hand, from the 2MASS and \emph{Spitzer} project databases, for pointings in the ISICE program.

\begin{deluxetable*}{ccccccccc}
\tabletypesize{\scriptsize}
\tablewidth{0pt}
\tablecaption{Calculated ice column densities of the H$_2$O, CO$_2$, and CO bands.\tablenotemark{a}\label{tbl-3}}
\tablehead{   
\colhead{Object} & \colhead{Type\tablenotemark{b}} &\colhead{N(H$_2$O)} & \multicolumn{2}{c}{N(CO$_2$)~$\times~10^{17}$ cm$^{-2}$} & \multicolumn{2}{c}{N(CO)~$\times~10^{17}$ cm$^{-2}$} & \multicolumn{2}{c}{N(Minor)~$\times~10^{17}$ cm$^{-2}$} \\
\colhead{} &\colhead{} & \colhead{$\times~10^{18}$ cm$^{-2}$} & \colhead{in CO} & \colhead{in H$_2$O} & \colhead{CO$_{rc}$} & \colhead{CO$_{mc}$} & \colhead{OCN$^-$} & \colhead{CO$_{gg}$}}
\startdata

1  & star (M5, 19.4)                 & 2.3 $\pm$~0.3 & $<$~0.6                        & $<$~0.6           & \nodata                    & \nodata                    & \nodata                   & \nodata\\ 
2  & YSO                        & 3.9 $\pm$~0.3 & 4.7 $\pm$~0.2              & 0.                      & 6.0 $\pm$~2.7 & 0.7 $\pm$~1.0 & 0.                            & 2.4 $\pm$~0.6\\ 
3  & YSO                         & 3.3 $\pm$~0.1 & 3.1 $\pm$~0.3              & 0.6 $\pm$~0.2 & 3.3 $\pm$~2.5 & 2.0 $\pm$~2.9 & 0.                            & 1.9 $\pm$~0.4\\ 
4  & YSO                         & 3.1 $\pm$~0.1 & 5.0 $\pm$~0.1              & 0.                      & 0.                              & 3.1 $\pm$~0.9 & 0.1 $\pm$~0.5 & 1.1 $\pm$~1.0\\ 
5  & YSO                         & 1.6 $\pm$~0.3 & 1.3 $\pm$~0.9              & 0.5 $\pm$~0.6 & 0.                              & $<$~1.5 & 0.                 & 0.\\
6  & star $\downarrow$ & $<$~4.5            & $<$~1.2                        & $<$~0.1           & 0.                      & $<$~0.6 & 0.                  & $<$~0.7\\
7  & YSO $\downarrow$ & $<$~1.4            & $<$~0.7                        & $<$~0.3          & 0. & 0. & $<$~0.2 & 0.\\
8  & YSO $\downarrow$ & $<$~0.8            & $<$~2.1                       & 0.                      & $<$~2.3 & 0. & $<$~0.1 & 0.\\   
9  & star $\downarrow$ & $<$~1.7            & \nodata                         & \nodata            & \nodata                    & \nodata                    & \nodata                     & \nodata\\
10& star (K0g, 6.7)                & 0.7 $\pm$~0.04 & 0.7 $\pm$~0.7            & 0.5 $\pm$~0.5  & 0.4 $\pm$~6.9 & 1.2 $\pm$~2.4 & 0.                               & 0.\\ 
11& star $\downarrow$ & $<$~1.1            & 0.                                 & $<$~0.3             & \nodata            & \nodata                     & \nodata                     & \nodata\\
12& star (K0, 19.8)                 & 0.9 $\pm$~0.03 & 0.                               & 0.5 $\pm$~0.1    & 0.                               & 0.                            & $<$~0.2  & 0.\\ 
13& star (K0, 23.4)                 & 1.4 $\pm$~0.05 & 1.2 $\pm$~0.5          & 0.3 $\pm$~0.4    & 2.9 $\pm$~6.9  & 2.3 $\pm$~7.1  & 0.                               & $<$~1.0\\ 
14& star (K0g, 41.2)                & 2.6 $\pm$~0.04 & 3.3 $\pm$~0.4          & 0.3 $\pm$~0.3   & 5.4 $\pm$~10.5 & 1.8 $\pm$~20.2 & 0.                             & 0.\\ 
15& star $\downarrow$ & $<$~0.6             & 0.                               & $<$~0.3             & \nodata          & \nodata                        & \nodata                   & \nodata\\
16& star $\downarrow$ & $<$~0.5              & $<$~1.0                   & 0.                        & 0.    & $<$~1.9& 0.                  & 0.\\ 
17& star $\downarrow$ & $<$~1.0             & 0.                              & $<$~0.7              & 0.         & $<$~2.6& $<$~1.1 & 0.\\ 
18& star $\downarrow$ & $<$~1.1            & $<$~0.9                    & 0.                         & $<$~8.9 & 0.        & 0.                                   & $<$~0.8\\ 
19& star $\downarrow$ & $<$~0.6            & 0.                               & $<$~0.6              & $<$~0.8 & $<$~0.8   & 0.    & $<$~3.5\\
20& star $\downarrow$ & $<$~1.0            & $<$~0.4                    & $<$~0.8               & 0. & $<$~1.8  & 0.                               & $<$~1.8\\ 
21& star (K0, 14.0)                 & 1.2 $\pm$~0.1 & 0.                                & 1.3 $\pm$~0.5     & \nodata                     & \nodata                      & \nodata                      & \nodata\\ 
22& star (K0g, 14.6)              & 1.0 $\pm$~0.1  & 0.                              & 0.9 $\pm$~0.1      & 0.                               & $<$~1.9 & 0.                               & 0.\\
23& star (K0g, 27.3)              & 1.5 $\pm$~0.02 & 2.1 $\pm$~0.1         & 0.4 $\pm$~0.1     & 1.6 $\pm$~2.4  & 1.7 $\pm$~0.5   & 0.                               & $<$~0.6 \\
24& star (K0, 12.8)                & 0.8 $\pm$~0.1 & 0.2 $\pm$~1.0          & 0.6 $\pm$~0.6      & \nodata                     & \nodata                       & \nodata                      & \nodata\\ 
25& star (K0, 19.2)               & 1.4 $\pm$~0.2  & 2.0 $\pm$~1.2           & 0.5 $\pm$~1.0     & 3.9 $\pm$~20.5 & 3.9 $\pm$~9.2  & 0.                                & $<$~0.1\\ 
26& YSO                     & 1.5 $\pm$~0.1   & 0.                              & 1.2 $\pm$~0.1      & 0.                                & 1.7 $\pm$~2.0 & $<$~0.9   & 0. \\ 
27& star (K0, 17.4)               & 1.0 $\pm$~0.1   & 2.0 $\pm$~0.6         & 0.4 $\pm$~0.5      & \nodata                     & \nodata                       & \nodata                      & \nodata\\
28& YSO                     & 2.1 $\pm$~0.2   & 2.7 $\pm$~1.4          & 0.1 $\pm$~1.1      & 0.                                & 0.                              & $<$~0.1    & 0. \\ 
29& star (K0g, 32.7)             & 1.8 $\pm$~0.04 & 1.5 $\pm$~0.2         & 0.7 $\pm$~0.2      & 2.7 $\pm$~4.3   & 2.2 $\pm$~0.6  & 0.                               & 0.4 $\pm$~0.7\\ 
30& star (G4g, 14.9)             & 1.0 $\pm$~0.1  & 1.3 $\pm$~0.8          & 0.5 $\pm$~0.7      & 0.                                & 4.4 $\pm$~46.4 & 0.                               & 0.\\ 
\enddata
\tablenotetext{a}{Column densities marked \nodata indicate that no fit was performed on the spectrum in the wavelength range surrounding that spectral feature. In general, this is because the spectrum was not extracted for this region - either the flux was below the extraction threshold of ARF, or the region was too confused (due to overlapping spectra) for ARF to fit that object. A value of 0 indicates that a fit was performed, and a value of 0 was returned. Only in the second case can it be concluded that this spectral feature is not present towards this object.}
\tablenotetext{b}{Objects from c2d target tables were categorised by the c2d team. For the objects taken from the 2MASS catalogue, the type was determined based on their J, H, K$_S$ photometry as described in \S~\ref{sec-classification}. In addition, for background stars, the spectral type and A$_V$, as calculated during baseline fitting (\S~\ref{sec-baseline-fitting}), is shown in parentheses. All object marked with $\downarrow$ gave upper limits only, due to confusion in the region of the object. As baseline fitting of these objects was made with a polynomial (\S~\ref{sec-baseline-fitting}), a stellar type could not be determined for these background stars.}
\end{deluxetable*}

\subsubsection{Dispersion modelling}

Individual spectra are extracted by transforming the source positions from the target files to the dispersed frames, accounting for a) the distortion in the telescope optics and b) the displacement of the first order dispersion relative to the direct light position by approximately 60 pixels in the dispersion direction. The dispersion is subsequently modelled using the linear relation of 0.0097~$\mu$m\,pixel$^{-1}$ with a vertical (non-dispersion, X, direction) correction of 0.00746929~pixel\,row$^{-1}$ to take into account the deviation of the peak position of the flux (as can be seen by inspection of Figure~\ref{fig-pipeline}a).

\subsubsection{Point spread function}\label{psf}

The PSF of the AKARI NG disperser was calculated by fitting various combinations of Gaussian functions or Lorentzian functions to a bright point source observed in AOT04 mode (Observation ID: 5020032\_001, standard star KF09T1, see \citet{Sakon12} for further details). The best fitting combination was a double Gaussian function:
\begin{eqnarray}\label{eqn-psf}
  f(X) =\left[ A_{1}\,exp\bigg\{-\left(\frac{X-P_{1}}{\sigma_1}\right)^2\bigg\}\right. + \nonumber \\
    \left.0.8 A_{1}\,exp\bigg\{-\left(\frac{X-\left(P_{1}+2.7\right)}{\sigma_2}\right)^2\bigg\} \right], 
\end{eqnarray}
where X is X pixel; $A_{1}$ the peak height and P$_{1}$ the peak position of the first Gaussian function; and $\sigma_n$ the full width half maximum (FWHM) of Gaussian functions (n = 1,2).

\subsubsection{Running the extractor}\label{sec-extraction-running}

For rows containing more than one target, a linear sum of PSFs is fitted when any targets are within five pixels of each other (in X). This is the key strength of the reduction method presented here, as this approach allows the separation of spectra which are overlapping in X, which would otherwise be discarded due to confusion. For each object in each row, the flux (in ADU) of that object at that point in its dispersion is then extracted by integrating the calculated PSF. In this way, the spectrum of each object is extracted systematically, row by row.

\begin{figure*}[htb]
\plotone{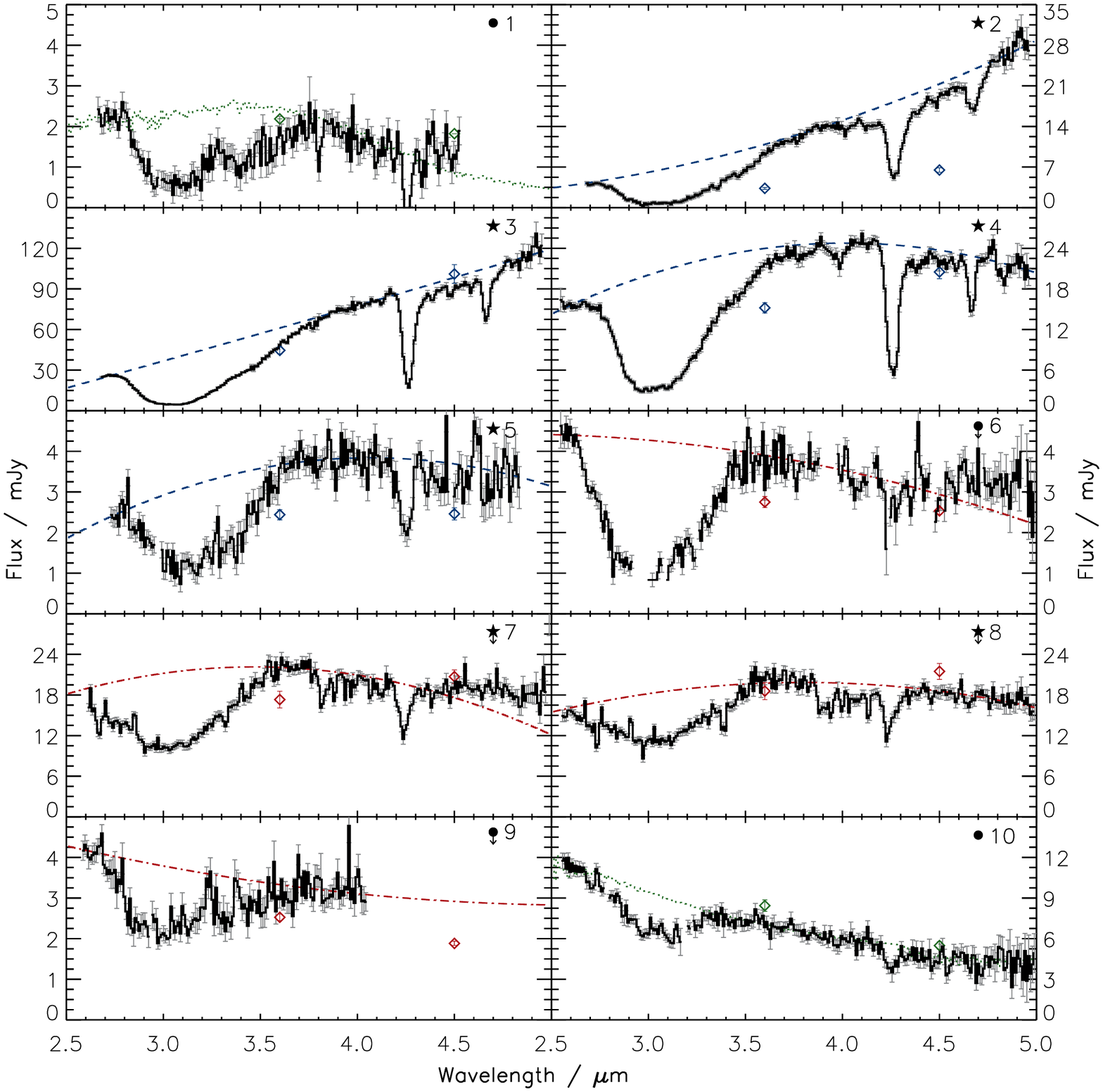}
  \caption{Spectra of all extracted objects on a flux scale, numbered as in Table~\ref{tbl-2}. YSOs are denoted by $\star$, while background stars are labelled $\bullet$. Where spectra provide only upper limits on molecular column densities, they are labelled with an arrow. IRAC photometric data is overplotted as diamonds. In most cases, the photometry matches well with the extracted spectra. Baselines were fitted to all of the spectra (see \S~\ref{sec-baseline-fitting} for full details) and these are overplotted with the following styles: (blue) dashed for YSOs, (green) dotted for background stars, and (red) dot-dashed for upper limits.\label{fig-spec1}}
\end{figure*}
\addtocounter{figure}{-1}

\subsection{Flux and wavelength calibration}\label{sec-wavelength}

The SRF was generated from an observation of standard star KF09T1 in the 1'~$\times$~1' aperture, with the pixel-to-wavelength relation determined by an observation of NGC 6543 in the Ns slit.
The spectrum extracted from this object was compared with its calibrated model template spectrum \citep{cohen99} and the response at each wavelength was obtained in units of ADU\,Jy$^{-1}$. The obtained SRF is consistent with that used in the latest IRC Spectroscopy Toolkit Version 20110114.

All extracted spectra are divided by the SRF to produce flux- and wavelength- corrected spectra, $F(\lambda)$ in units of millijansky (mJy), as illustrated in Figure~\ref{fig-pipeline}d. The final spectrum of an object which appears in two pointings is generated by combining both spectra after division by the SRF; the two spectra are summed, and the flux divided by two. For all spectra produced this way (not including upper limits, see \S~\ref{sec-results}), the difference between the two summed spectra was less than 5~\%, which was taken into account in the calculated error on the final, combined spectrum. 

\subsection{Error propagation}\label{sec-error}

The flux error in each pixel is $\pm$4~ADU (Youichi Ohyama, \emph{Private communication}). Thus the minimum error in each stacked pixel, $\epsilon_{min}$, is calculated by:

\begin{equation}\label{eqn-stack_err}
\epsilon_{min} = \sqrt{n \sqrt{2 \cdot 4^2}},
\end{equation}
where n is the number of frames stacked, and the factor of two accounts for a contribution from subtracting the dark from each frame prior to stacking. During stacking, the standard deviation in the flux, $\epsilon_{sd}$, is calculated. These errors are combined to calculate $\epsilon_{ADU}$, the overall error in flux, used to weight the fit of the PSF during extraction. The error in the extracted flux, $\epsilon_{ext}$, is the combination of the data error $\epsilon_{ADU}$ with the standard deviation of the fit. During division by the SRF, the final error on each datapoint, $\epsilon$, is calculated by combination of the extracted flux error with the error on the SRF.

\section{Observational results}\label{sec-results}

\begin{figure*}[htb]
\plotone{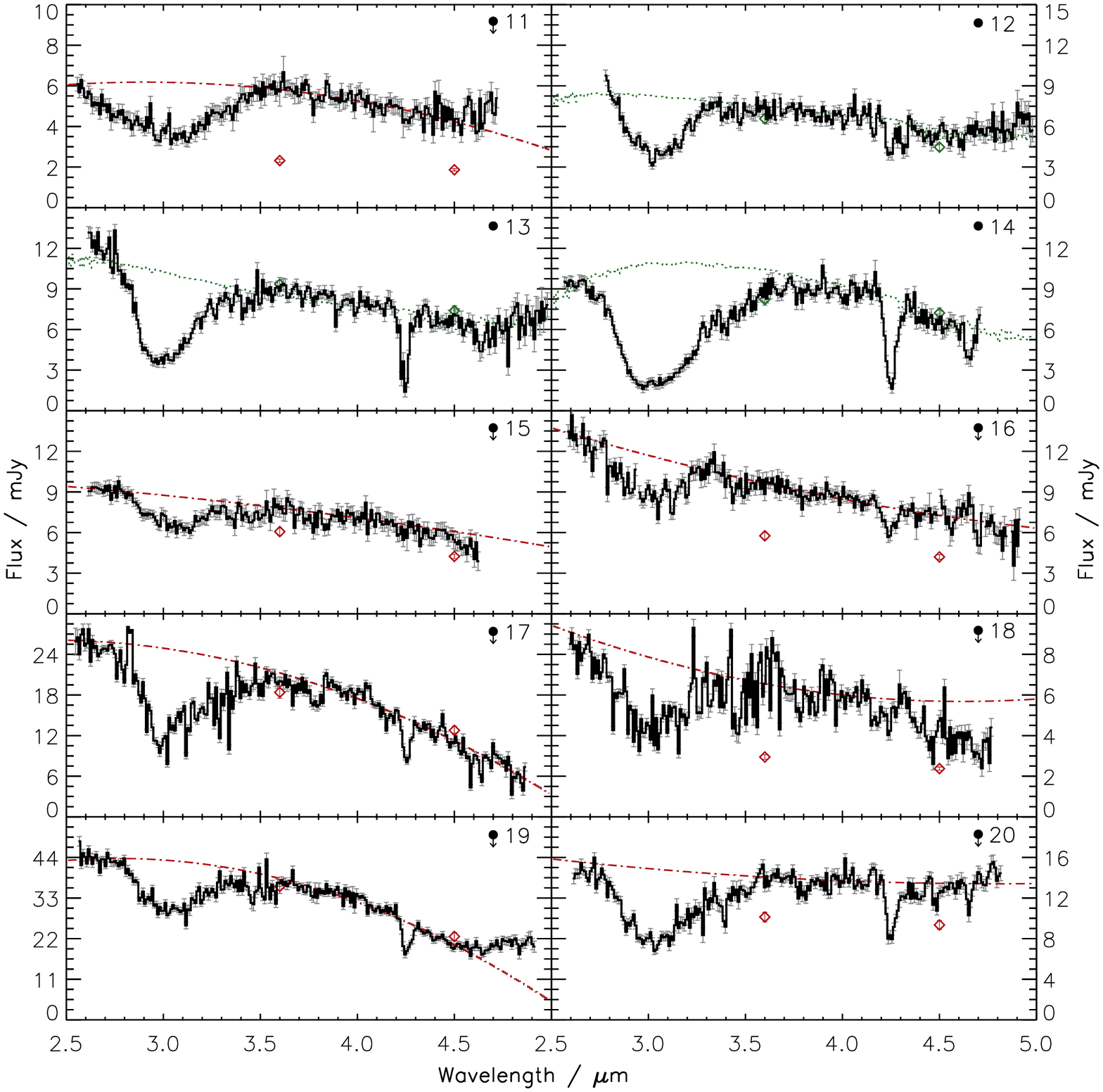}
\caption{continued.\label{fig-spec2}}
\end{figure*}

In total, across the 19 clouds there were 94 target objects with some 2MASS K$_S$ band magnitude. 27 of these objects fell below the signal-to-noise threshold of ARF, defined as K$_S$ = 1.19~mJy, the magnitude of the brightest isolated object for which the PSF fitting failed to converge (including all objects in DC~269.4+03.0 and L~1082A.) The confusion limit for galactic cirrus and point sources is $<$~1~$\mu$Jy in N3, the 3~$\mu$m imaging filter \citep{ast05}. Four objects were only partially observed due to their position at the edge of the field of view, and were thus discounted. 33 objects were too confused to extract any spectral data because of their low flux and overlap with nearby objects, including all those in DC 300.2-03.5. The objects which were successfully extracted using ARF are detailed in Table~\ref{tbl-2}; those derived from only a single spectrum are denoted by a dagger symbol ($\vdag$). The spectra of the 30 extracted objects are presented in Figure~\ref{fig-spec1}. The final spectra are shown in black, with the error on each point overplotted in grey. 

Spectra were extracted across the full 2.5~--~5~$\mu$m region, except in cases where the dispersion reached the edge of the detector or where the flux from two neighbouring objects could not be resolved.  For example, due to detector limitations, in Object~9 the region of the spectrum beyond 4~$\mu$m was not obtained, while for a further four objects (Objects 1, 21, 24 and 27), the higher wavelength region of the spectrum, beyond 4.5~$\mu$m, was not extracted. Where unresolved overlap occurred between objects, it was possible to extract a partial spectrum, or a spectrum with some extra flux contribution, from which upper limits could be obtained on the molecular species present in the line of sight (Objects 6, 7, 8, 9, 11, 15, 16, 17, 18, 19 and 20). These objects are indicated by an arrow in Table~\ref{tbl-3} and are designated as upper limits in all figures ($\downarrow$). 

The H$_2$O band (3.0~$\mu$m) was fully extracted for all objects, including the blue wing, which is not observable from the ground. In many spectra, such as Objects~2 and 4, it is clear that the H$_2$O band is saturated. The error values calculated in the pipeline are the errors on the observed flux, not the errors on the absolute value of the spectral data. Where we suspect bands are saturated, we have not included any additional uncertainty on these data points. In at least 12 objects the water band is still clearly visible even though the flux is $<$~5~mJy (see Figure~\ref{fig-spec1}, Objects~1, 5, 6, 9, 21, 22, 24, 25, 26, 27, 28, 30). The absorption bands of CO$_2$ (4.25~$\mu$m) and CO (4.67~$\mu$m) are also clearly visible for the majority of objects.

\addtocounter{figure}{-1}
\begin{figure*}[htb]
\plotone{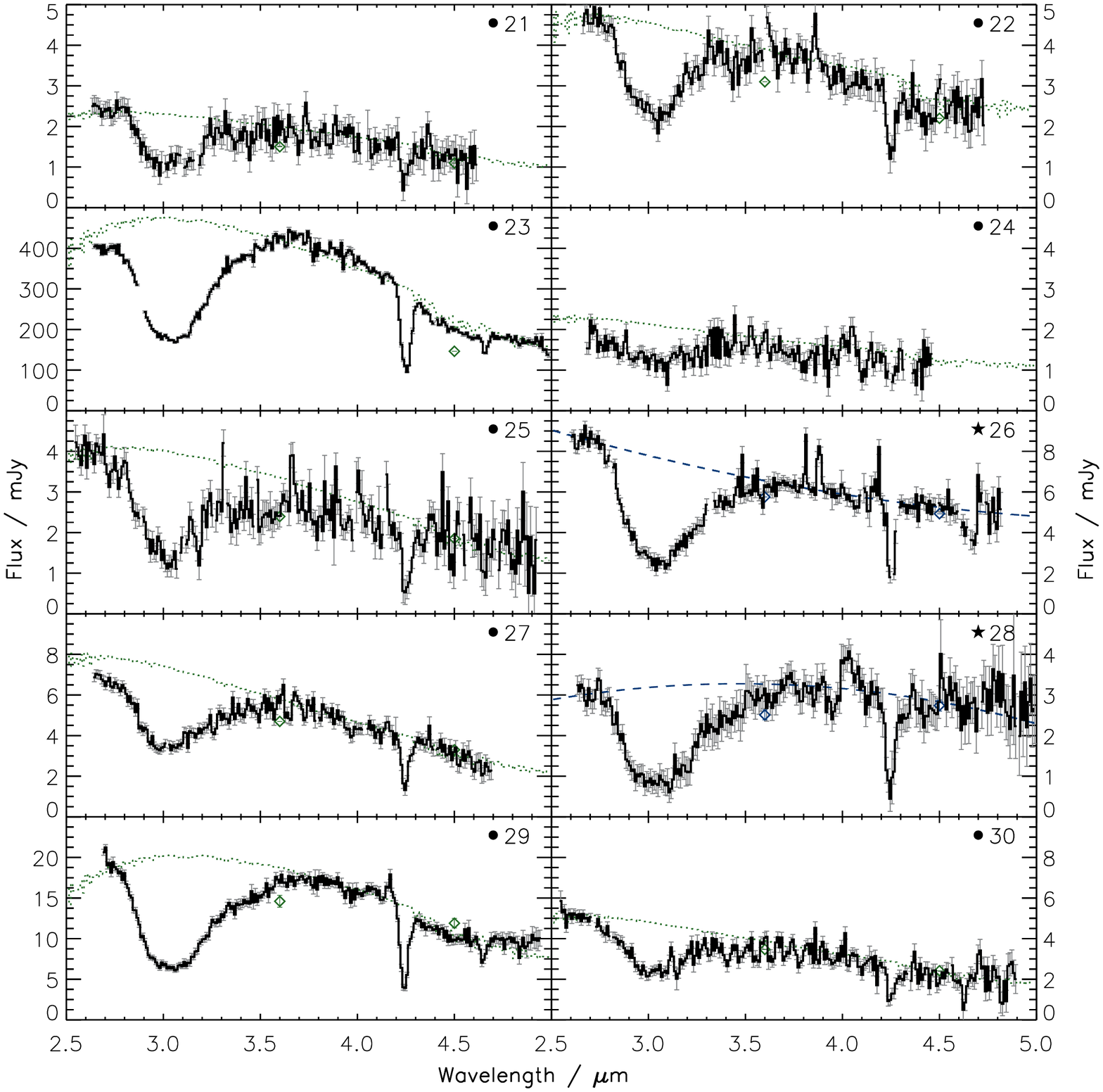}
  \caption{continued.\label{fig-spec3}}
\end{figure*}

On each spectrum in Figure~\ref{fig-spec1}, the IRAC photometric data points are plotted as diamonds at 3.6 and 4.5~$\mu$m. For most objects the photometric data matches the extracted flux very well. For some of the upper limits mentioned above, notably Objects~9, 11 and 16, the IRAC photometry does not exactly match the flux of the object, suggesting that there was excess light on the detector in the vicinity of the spectrum, vindicating the decision to only quote upper limits to the molecular column densities in these objects. In all cases where there is a slight photometry mismatch, the flux of the object has been over-, rather than underestimated, supporting the conjecture that there is excess light in these fields of view. The only two examples of completely isolated objects in the dataset are Objects~27 and 30. The photometric data points for these two objects match the extracted flux within the errors.

\subsection{Object classification}\label{sec-classification}

For further spectral analysis, it was necessary to classify each object according to whether it was a field star behind the molecular cloud (``star'') or a YSO embedded in the cloud (``YSO''). All objects in the IMAPE programme had a classification of this type preassigned by the c2d team, based on the full \emph{Spitzer} photometric data as well as 2MASS photometry. The classification of objects in the ISICE programme was based on 2MASS photometric data. If objects satisfied: 
\begin{equation}
  1.75(H-K_S)-0.04 \leq J-H,
\end{equation} \label{eqn-colour-pick}       
they were classed as background stars; if not, they were classed as YSOs \citep{ito96,she08}. 
Henceforth, irrespective of the observational programme from which they are derived, the objects in this paper are designated by a star symbol ($\star$) for YSOs, or a circle ($\bullet$) for background stars; in the text they are referred to as ``YSO'' and ``star'', respectively. 

\subsection{Baseline fitting}\label{sec-baseline-fitting}

To convert the observational data to spectra on an optical depth scale (most relevant for molecular abundance analysis), two different methods of baseline fitting were employed. As is typical in ice spectra extraction \citep{gib04,Oberg11,Aikawa12}, this was either a second order polynomial (in the case of YSO objects) or a NextGen model \citep{hauschildt99} (in the case of background stars), with the proviso that second order polynomials were also fitted to ``upper limit'' spectra, i.e. where the data were not fully extractable. These baselines are shown overplotted on each individual spectrum in Figure~\ref{fig-spec1}, as either a (green) dotted line (NextGen models, background stars), a (blue) dashed line (second order polynomials, YSOs) or a (red) dash-dotted line (second order polynomials, ``upper limit'' spectra). 

The NextGen models were fitted to the background stars based on photometric data from 2MASS, IRAC and MIPS observations. Using the extinction law of \citet{wei01}, objects were dereddened along a reddening vector of 2.078 on a (J-H) vs (H-K$_S$) colour-colour diagram to determine their spectral type and A$_V$ (tabulated in Table~\ref{tbl-3}). Potential spectral types were defined based on where the dereddened object crossed the main sequence, giant or dwarf branches.
This approach accounted well for the reddening in the 2MASS wavelength region of the spectra, but at longer wavelengths the model baselines were further reddened by a polynomial.

The advantage of fitting baselines with NextGen models is that the models account for emission and absorption in the stellar atmosphere. However, these models can not be used to fit YSOs as the NextGen models only include main sequence and giant stars. The continua of YSOs, (and all upper limits on background stars and YSOs) were therefore fitted by second order polynomials. Fitting regions varied slightly between the spectra, based on the position of bad pixels, but were generally $<$~2.65~$\mu$m, 3.85 -- 4.18~$\mu$m, and $>~$4.75~$\mu$m. The fits were weighted in favour of the blue end of the spectrum, as this region had a higher signal-to-noise ratio than the red end. 

Spectra were converted to optical depth scale by:
\begin{equation}\label{eqn-od}
\tau =-ln(F/F_{con}),
\end{equation}
where $\tau$ is optical depth, $F$ is the spectral flux, and $F_{con}$ the flux of the continuum (fitted in \S~\ref{sec-baseline-fitting}). The optical depth scale spectral features are shown in Figure~\ref{fig-od-1a} for the 3~$\mu$m H$_2$O band, Figure~\ref{fig-od-3a} for the 4.67~$\mu$m CO band, and Figure~\ref{fig-od-2a} for the 4.25~$\mu$m CO$_2$ band.

\section{Spectral analysis and discussion}\label{sec-colden}

The column densities of molecular species can be determined by integrating the area under their absorption bands on an optical depth scale and correcting by the optical constant for each species, using: 
\begin{equation}\label{eqn:colden}
 N=\int \frac{\tau d\nu}{A},
\end{equation} 
where $N$ is column density in molecules~cm$^{-2}$, $\nu$ is wavenumber in cm$^{-1}$, and $A$ is the band strength in cm molecule$^{-1}$ \citep{gerakines95}. The specific analysis methods for H$_2$O, CO$_2$, and CO are described in detail below. All calculated column density values are presented in Table~\ref{tbl-3}. 

The absorption bands of interstellar ice species are altered, with respect to those measured in the laboratory, by interaction of the incident light with the dust grains on which the absorbing ices are present. This is particularly important in the NIR where the average grain radii and absorption wavelengths are similar. Therefore for an abundant ice species, if a laboratory spectrum is to be used to derive the column density of the ice by fitting it to the observational data, a grain shape correction must be applied. In this work, a continuous distribution of ellipsoids (CDE) model was employed. CDE is a reliable model of fractal aggregates (the likely structure of interstellar grains) and it has been successfully applied to many astronomical spectra \citep{tie91,ehr97,pon03}.

Due to the asymmetric shape of the PSF of AKARI IRC data (described above in \S~\ref{psf}), the instrumental line profile was determined by analysis of an observation of the standard star KF09T1 in the imaging mode (IRC03). The imaging frame N3 (3.2~$\mu$m) \citep[see ][]{ona07} was chosen, and the central four pixels in the X direction were summed to produce the profile of a PSF along the Y direction, which was assumed to be the instrumental line profile.


\begin{figure*} 
\plotone{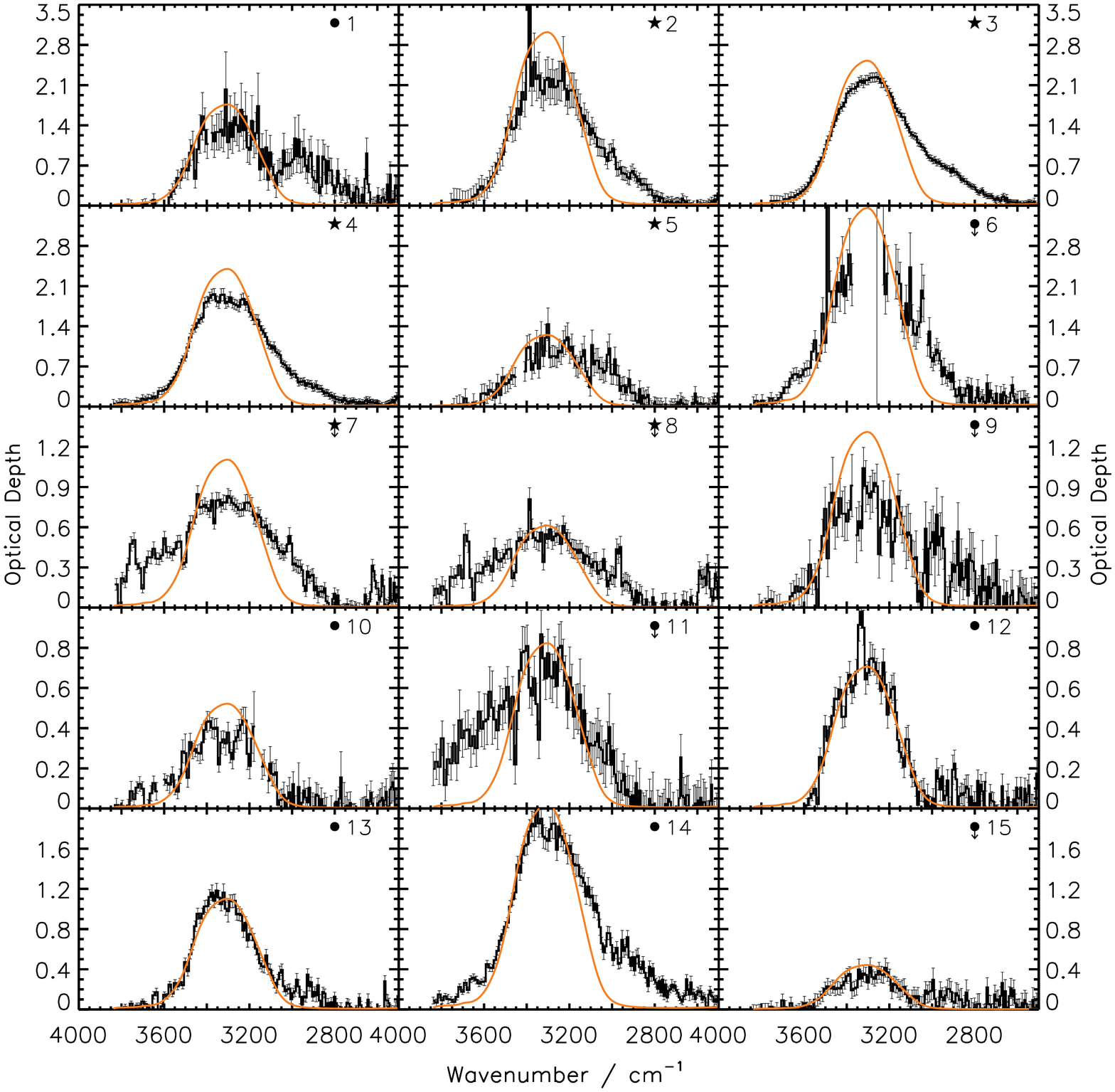}
  \caption{The fits of a pure non-porous H$_2$O laboratory spectrum at 15~K, corrected with a CDE model, the instrumental line profile, and convolved to the resolution of AKARI, to all H$_2$O absorption bands at 3~$\mu$m (3333~cm$^{-1}$). Observational data, on an optical depth scale, is plotted in black, with the fits overplotted in orange.\label{fig-od-1a}}
\end{figure*}
\addtocounter{figure}{-1}
\begin{figure*} 
\plotone{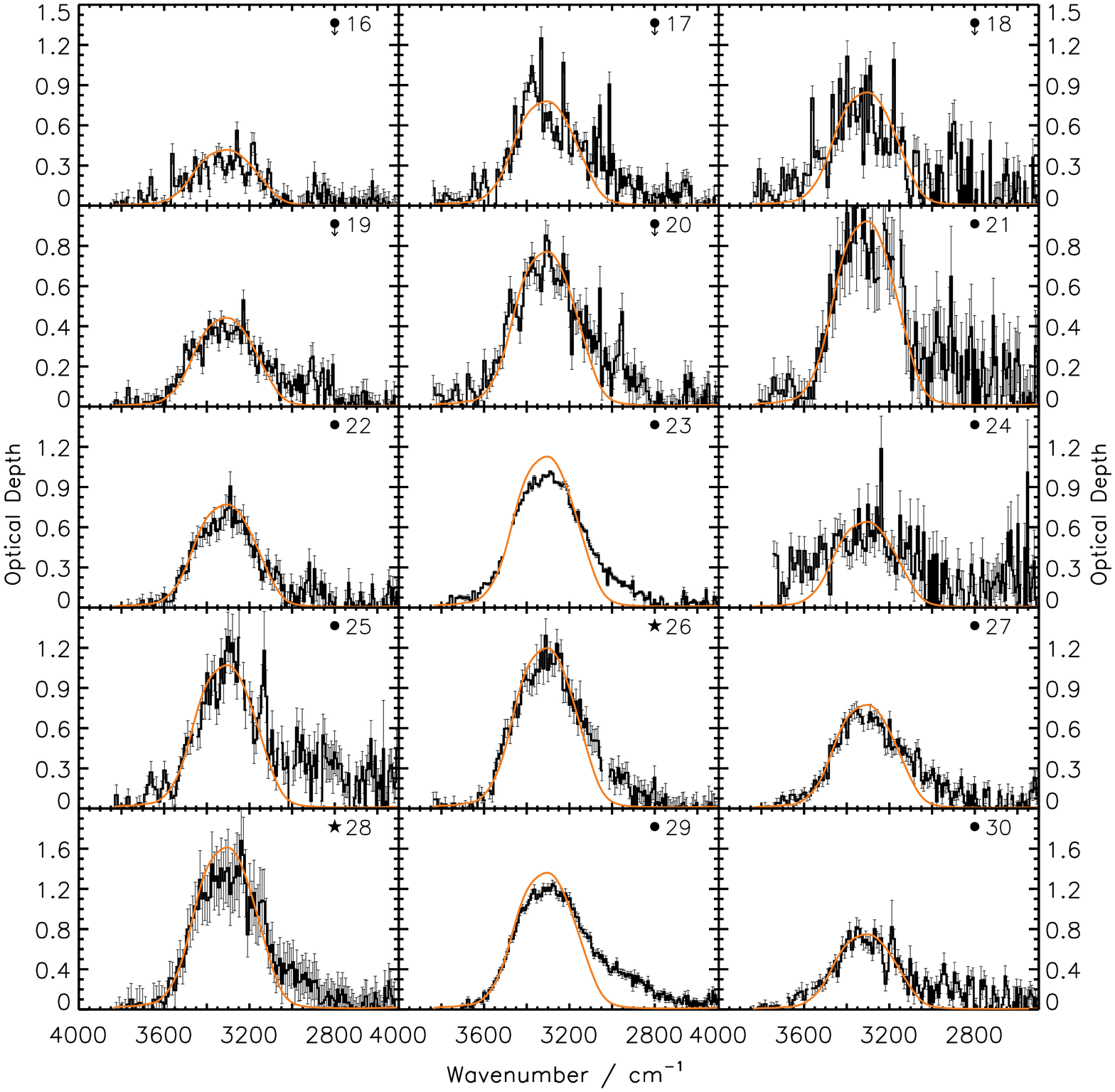}
  \caption{continued.\label{fig-od-1b}}
\end{figure*}

\begin{figure*} 
\plotone{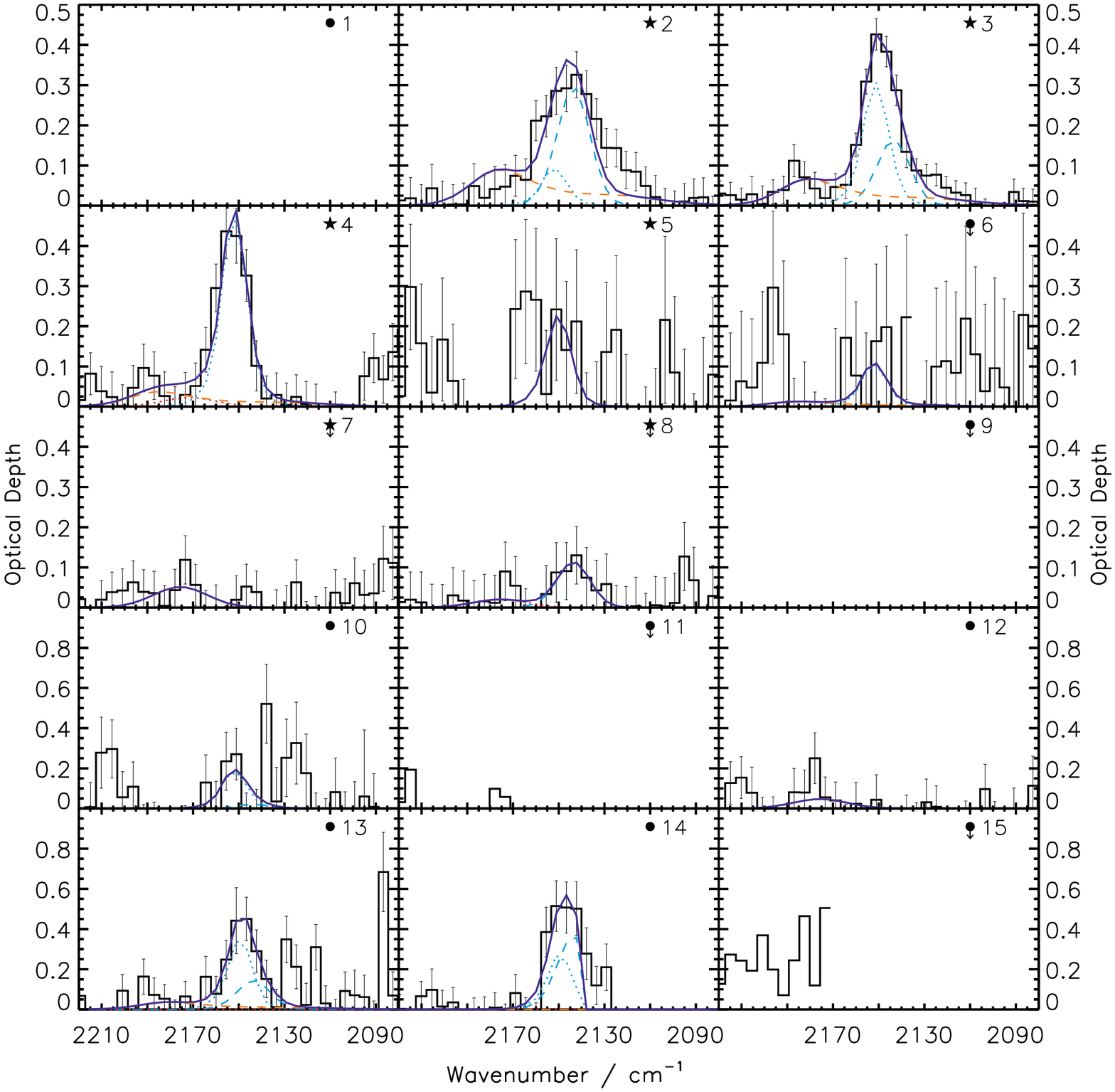} 
  \caption{Fits to all CO absorption bands using a component analysis method. Observational data, on an optical depth scale, is plotted in black, with the fits overplotted as follows: CO$_{rc}$ light blue dashed, CO$_{mc}$ light blue dotted, OCN$^-$ red dotted, CO$_{gg}$ orange dashed, and the overall fit in dark blue (solid). CO$_{mc}$ is CDE-corrected to account for grain shape effects, and the sum of all components is corrected for the AKARI instrumental line profile and resolution. Where there is no data in a plot, it is because no data was extracted in this region of the spectrum e.g. Object~1. Where there is only a partial data extraction, the plot is shown, but no fit was made to the data e.g. Object~11.\label{fig-od-3a}}
\end{figure*}
\addtocounter{figure}{-1}
\begin{figure*} 
\plotone{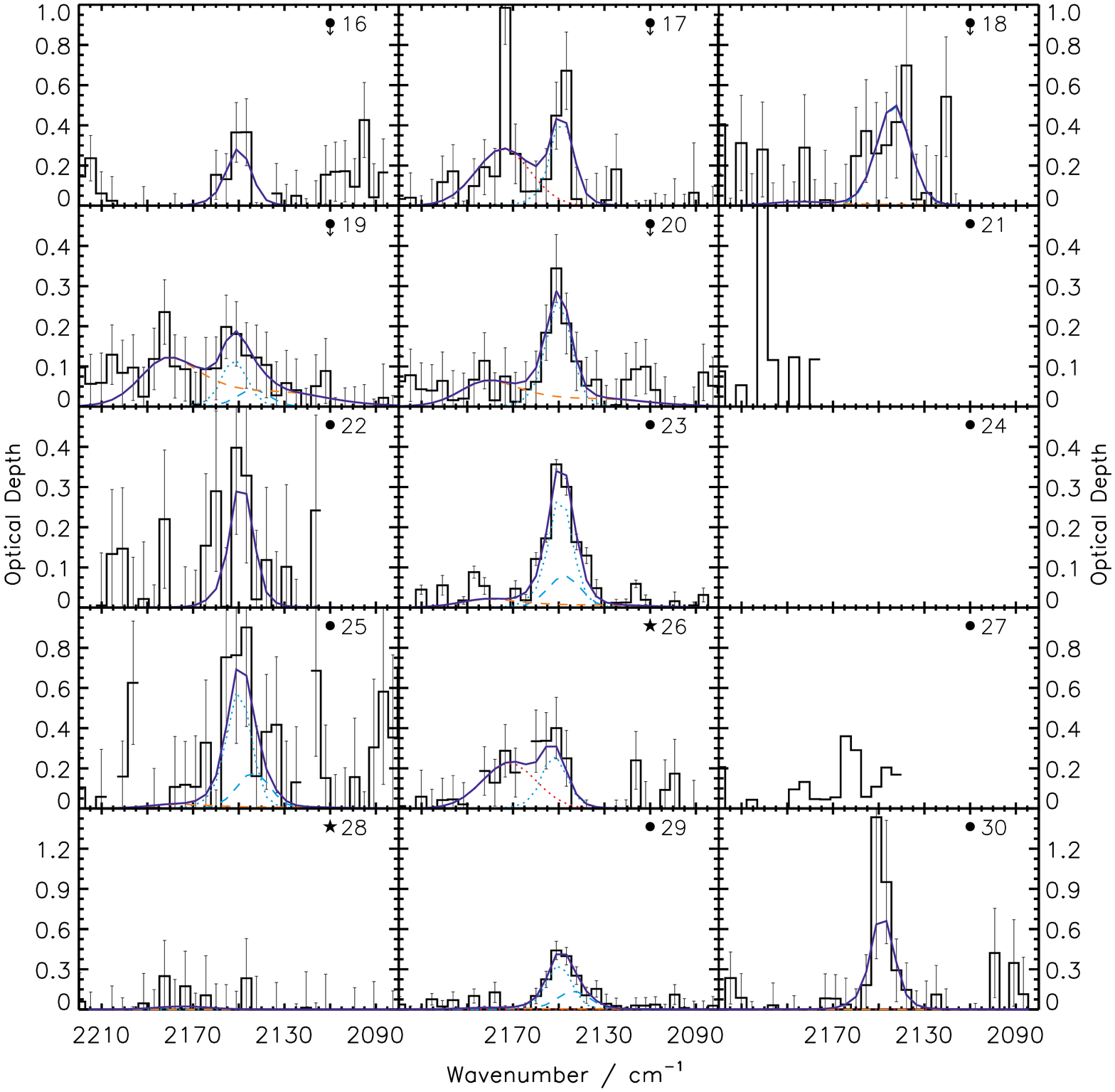}
  \caption{continued.\label{fig-od-3b}}
\end{figure*}

\begin{figure*} 
\plotone{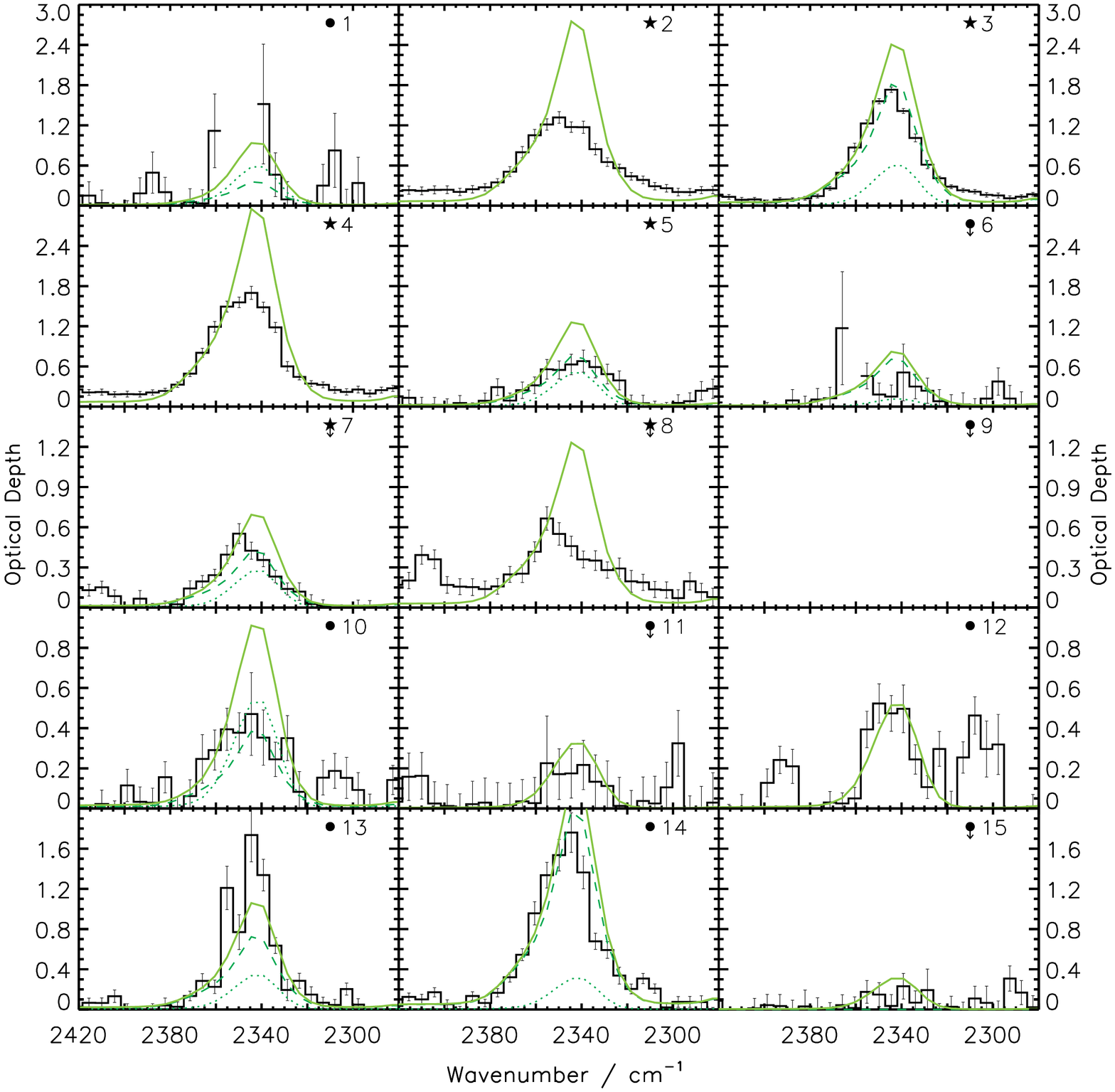} 
  \caption{Fits to all CO$_2$ absorption bands using a combination of H$_2$O-rich and CO-rich CO$_2$ laboratory ices. Observational data, on an optical depth scale, is plotted in black, with the fits overplotted as follows: CO$_2$ in a CO-rich ice is dark green dashed, CO$_2$ in a H$_2$O-rich ice is dark green dotted, and the overall fit in light green (solid). Both components are corrected for the AKARI instrumental line profile and resolution, and CDE corrected to account for grain shape effects.\label{fig-od-2a}}
\end{figure*}
\addtocounter{figure}{-1}
\begin{figure*} 
\plotone{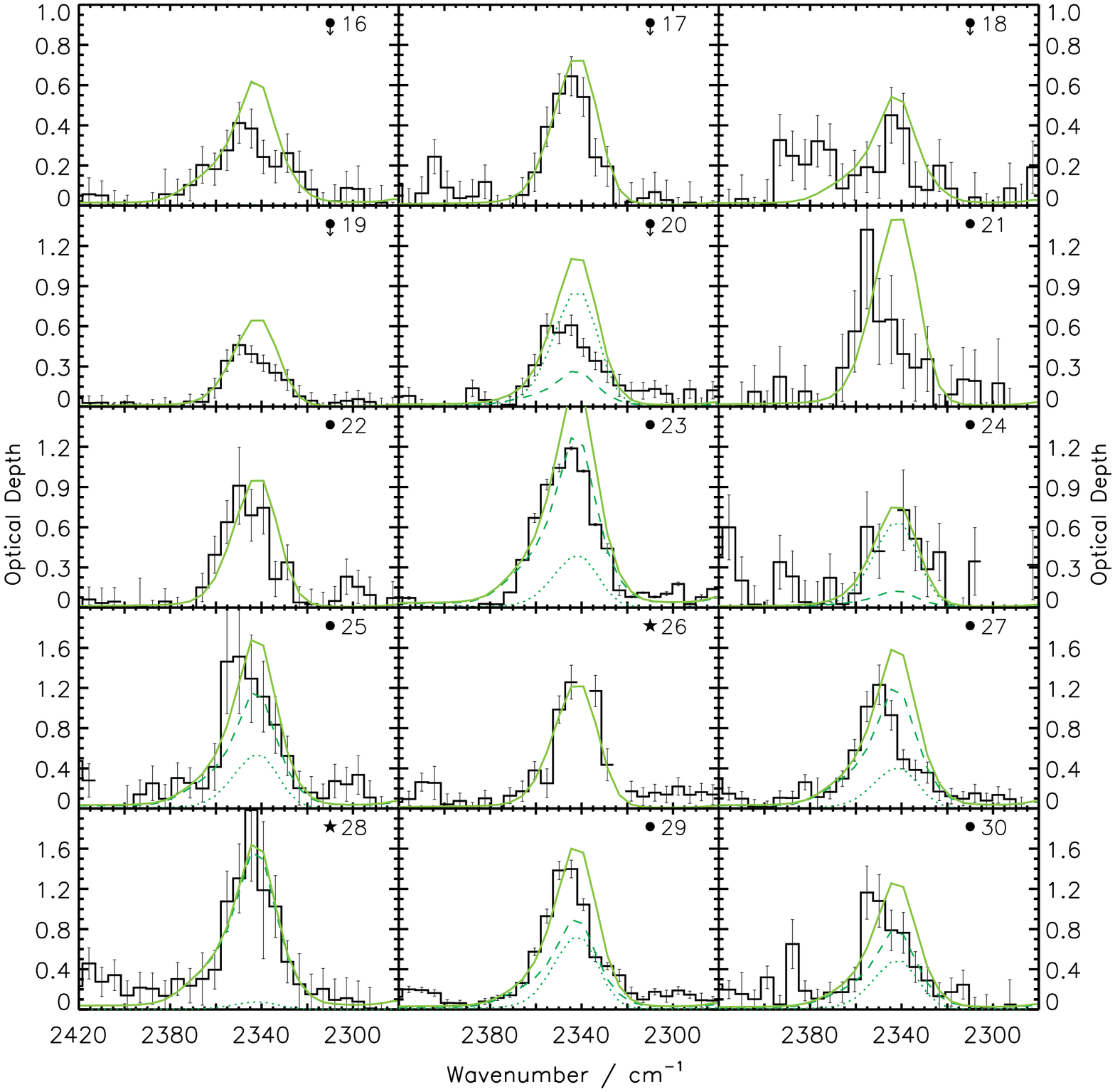}
  \caption{continued.\label{fig-od-2b}}
\end{figure*}

\clearpage

\subsection{H$_2$O}\label{sec-h2o}

H$_2$O was detected in all 30 extracted spectra, as shown in Figure~\ref{fig-od-1a}. These spectra include a broad, smooth band extending from $\sim$ 3600 to 2600 cm$^{-1}$ with a steep blue wing and, in almost all cases, an extended red wing. In order to fit this H$_2$O absorption feature at $\sim$~3~$\mu$m (3333~cm$^{-1}$), a CDE-corrected spectrum of H$_2$O was generated from a laboratory spectrum of pure H$_2$O at 15~K \citep{Fraser04}, then corrected for the instrumental line profile and resolution of AKARI. This spectrum was fitted exclusively to the blue wing of the observed H$_2$O data, from approximately 3600~--~3400~cm$^{-1}$. Unlike the red wing, which is discussed in detail below, this region is unaffected by grain interaction effects, and is accessible here for the first time since ISO. The blue wing is also unaffected by saturation. As can be seen in Figure~\ref{fig-od-1a}, the pure H$_2$O spectrum fitted the H$_2$O bands well in all cases. 

Crystalline H$_2$O has a characteristic peak shape and position, and has been observed in the circumstellar disks around YSOs, where material is subject to intense irradiation from the newly formed star \citep{Malfait98,Honda09}. For all objects, test fits were made to examine the possibility of a crystalline H$_2$O component towards each line of sight, but no crystalline ice was found. A single, amorphous H$_2$O lab spectrum at 15~K was used to fit every H$_2$O absorption feature in the 30 AKARI spectra. Column densities of H$_2$O were calculated from the fitted laboratory spectrum using Equation~\ref{eqn:colden}, with A~=~2~$\times$~10$^{-16}$~cm~molecule$^{-1}$ \citep{gerakines95}, and are presented in Table~\ref{tbl-3}.

\begin{figure}[htb]
  \plotone{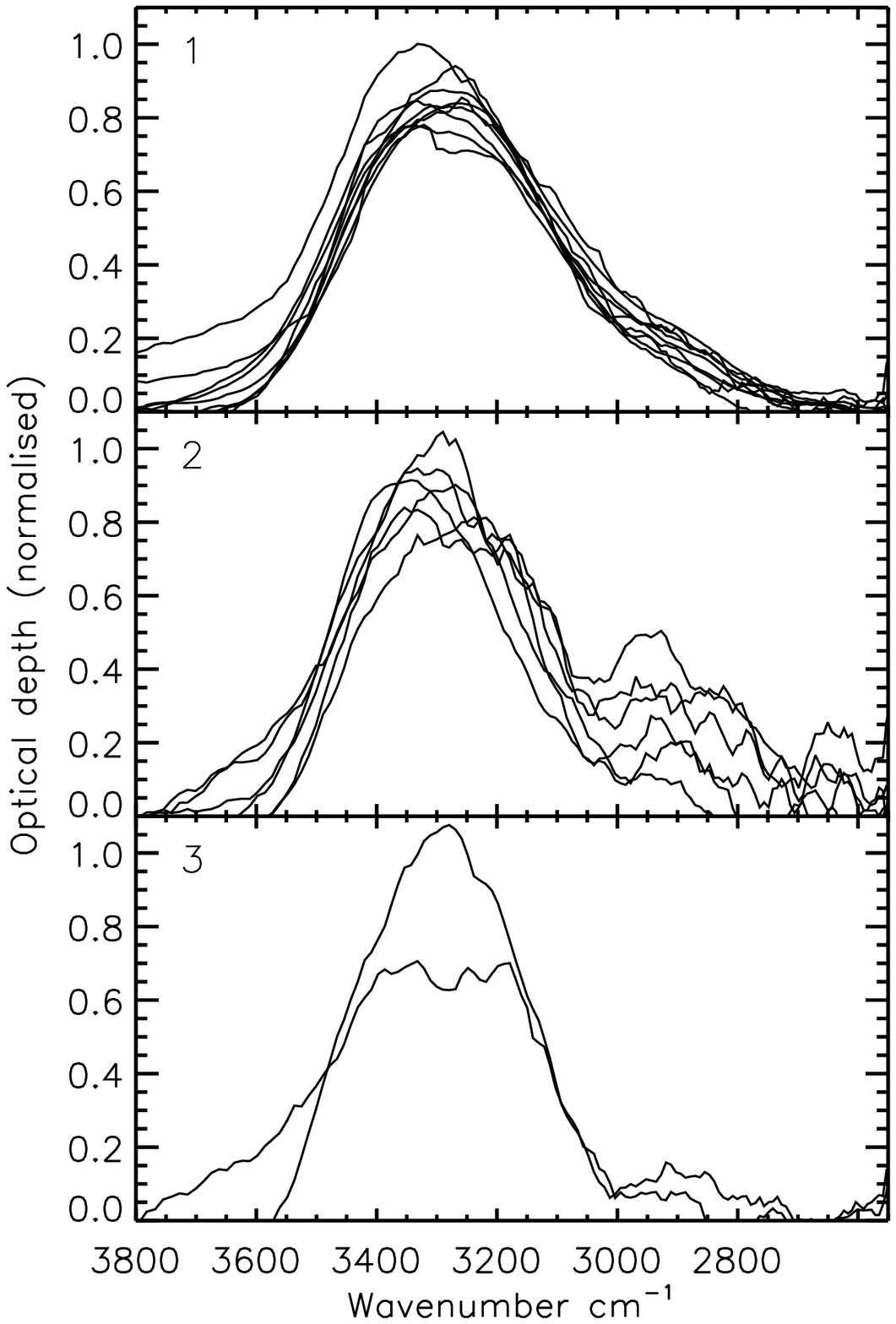}
\caption{H$_2$O absorption band profiles for Types 1, 2, and 3. These data have been smoothed, normalised and plotted without errors, to aid comparison. It is particularly evident in this figure, from the flattened peak of the band profile, that the H$_2$O band is almost always saturated.}\label{fig-h2o-bands}
\end{figure}

Excluding those spectra yielding only upper limits, the H$_2$O profiles of all objects can be divided into three types, as illustrated in Figure~\ref{fig-h2o-bands}. In this figure, all spectra are normalised to the maximum of the fitted H$_2$O band at 3300~cm$^{-1}$. The three distinct H$_2$O profiles identified were: Type 1, in which the red wing is weak compared to the absorption peak at 3~$\mu$m; Type 2, in which the long wavelength feature is relatively strong, i.e. a separate shoulder; and Type 3, containing no (or very little) red contribution at all. Similar effects were noted by \citet{Thi06}, who found two H$_2$O band profiles in the spectra of intermediate mass YSOs in Vela: one where the extended red wing was weak relative to the H$_2$O ice band, and a second where it was stronger. As in \citet{Thi06}, within each subset of objects the H$_2$O bands show remarkably similar profiles. The question is, what are the likely origins of these profile differences? Do these profiles represent distinct physical environments, or have we, rather, sampled a continuum of red wing profiles?

Type 3 is seen only in the spectra of Objects 10 and 12, both of which are background stars and have low spectral flux ($<$~12~mJy at all wavelengths observed). No previous observations exist of H$_2$O bands without an extended red wing emission. For these bands, a CDE-corrected pure H$_2$O spectrum appears to fully fit the observed data. Type 2 -- present in the spectra of Objects~1, 13, 21, 22, 25 and 30, a sample of only background stars -- appears to contain a distinct red shoulder, as opposed to the usual extended red wing. Type 1 -- present in the spectra of Objects~2, 3, 4, 5, 14, 23, 24, 26, 27, 28 and 29, a mixture of YSOs and background stars -- resembles the ``standard'' water band observed towards many lines of sight \citep{gib04,Boogert11}. The origins of the extended long wavelength feature on the water ice band have often been debated. A number of suggestions have been made as to its origin, including: the interaction of light with dust grains (as discussed in \S~\ref{sec-colden}), with the depth and profile of the band changing with grain size, shape, or even as a function of ice mantle thickness \citep{smith1989}; an absorption band at 3.54~$\mu$m, attributed to the C-H stretch of methanol, CH$_3$OH \citep{hagen80,dar03}; and aliphatic and aromatic C-H stretches associated with various molecular sources, including polyaromatic hydrocarbons \citep{Onaka11}, larger organic molecules \citep{Brooke99} and the grains themselves \citep{jones12a}.

\begin{figure}
   \plotone{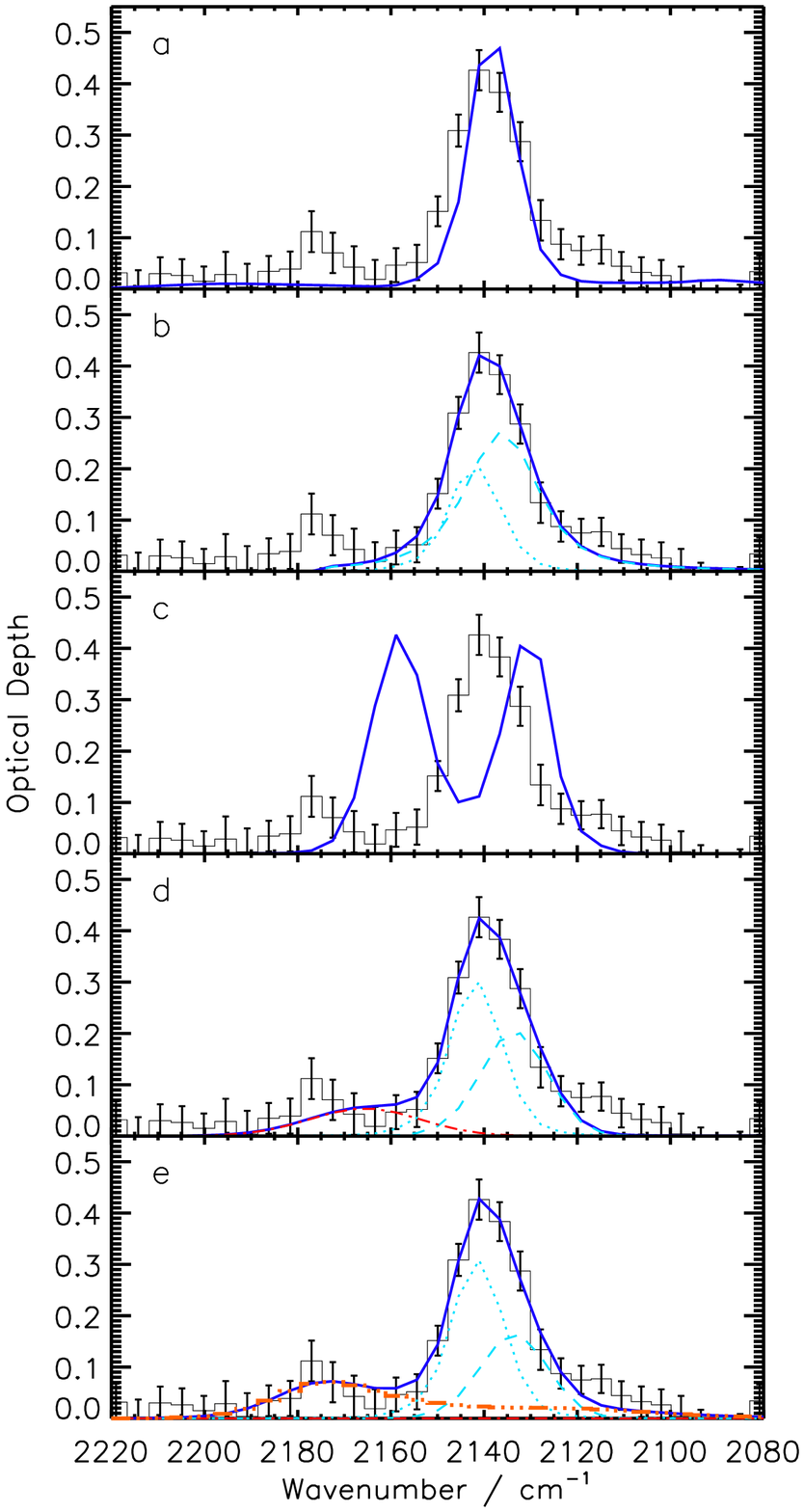} 
\caption{The fitting methodology for the CO absorption feature, illustrated for Object~3. Various methods were attempted before picking the final, best-fitting approach. a) a pure CO laboratory spectrum \citep{Ehrenfreund96}; b) a combination of CO$_{rc}$ (dotted, blue) and CO$_{mc}$ (dashed, blue) \citep{pon03}; c) a modelled CO gas phase absorption at 15~K; d) a combination of CO$_{rc}$, CO$_{mc}$, and OCN$^-$ (dot-dashed, red) \citep{pon03,vanb05}; e) the final fitting approach, with a combination of CO$_{rc}$, CO$_{mc}$, OCN$^-$, and CO$_{gg}$ (dot-dot-dashed, orange) \citep{pon03,vanb05,fraser05}.}\label{fig-COgg1}
\end{figure}

\citet{Thi06} suggest that, although their sample is statistically limited, a strong correlation is evident between the water ice column density and the optical depth of the extended red wing at 3.25~$\mu$m, concluding that the carriers of the red wing are directly related to the ice mantle. A previous study by \citet{Smith93} also finds a strong correlation between Av and the optical depth of the extended red wing at 3.47~$\mu$m. Neither of these correlations is evident in our data.
However, in comparing the average H$_2$O ice column densities across Types 1, 2, and 3, a clear trend emerges: the column density of H$_2$O ice in Type 3 sources is $\sim$~2.6 times lower than Type 1; Type 2 is $\sim$~1.5 times lower than Type 1. Given that Types 3 and 2 only contain background stars whereas Type 1 also includes YSOs it appears that the profile and depth of the extended red wing could potentially prove to be a reliable evolutionary marker. Within Type 1, whilst the band profiles are identical, the average H$_2$O ice column densities differ by a factor of 1.7 between the subset of YSO objects and the subset of background stars. In comparison, the ratio of average column densities between the YSO subset in Type 1 and the Type 2 background stars is 1.9. The objects of \citet{Thi06} were all class I YSOs; the average H$_2$O column densities associated with their Type 1 profile (i.e. the ``standard'' water band) is identical to ours (2.5~$\pm$~0.9 $\times$~10$^{18}$, and 2.1~$\pm$~0.5 $\times$~10$^{18}$ molecules cm$^{-2}$, respectively). This provides statistical evidence suggesting that the H$_2$O band profile also changes with the column density of ice in a particular line of sight.

It has previously been observed that the water ice abundance (H$_2$O/H$_2$) in molecular cores increases with dust density at low A$_V$ \citep{Whittet88}, before reaching a plateau \citep{Pont04}. Some evidence exists of a further jump in water ice abundance at densities of a few $\times$~10$^5$~cm$^{-3}$ H$_2$, which has been attributed to the removal of one or more water destruction pathways \citep{Pont04}. From our calculated column densities, it is likely that the YSOs observed in our survey probe lines of sight with densities near to the plateau, i.e. regions where grain growth and accelerated freeze-out have occurred, whereas the lines of sight towards background stars probe the density-abundance slope at lower densities, similarly to previous observations \citep[e.g.][]{Whittet88}. On the evolutionary path from background stars (probing low density molecular clouds or pre-stellar cores) to class I low-mass YSOs, it appears that both the grain radii and the ice mantle thickness increase. However, previous attempts to fully account for the extended red wing with this postulate have failed \citep[e.g.][]{smith1989,Thi06}. Observational evidence from other ice species also points towards fundamental differences in the dust present in diffuse and dense ISM regions \citep{Boogert11}. 

As mentioned above, the H$_2$O band of Type 3 objects is fully fitted with a CDE-corrected H$_2$O ice profile, suggesting that only H$_2$O contributes to this absorption feature. The column densities of H$_2$O towards these lines of sight are relatively low, with values of 0.7 $\times$~10$^{18}$ and 0.9 $\times$~10$^{18}$ molecules cm$^{-2}$ for Objects 10 and 12, respectively. In Type 2 objects, the main H$_2$O band centred at $\sim$ 3~$\mu$m is equally well accounted for by a CDE model as the H$_2$O observed towards Type 3 objects, but the column densities of H$_2$O are, on average, 1.8 times greater. The water ice band does not have an extended red wing, but rather a distinct shoulder redward of the H$_2$O band, attributed to additional molecular species present in the ice. Type 2 objects thus likely probe lines of sight towards more evolved regions of molecular cores where surface chemistry has resulted in the formation of species such as CH$_3$OH, or the hydrogenation of carbon on the grain surface, producing C-H bonds which absorb in this wavelength region. One further alternative, not widely discussed in the literature, is that on such lines of sight the nitrogen chemistry has evolved, forming NH$_3$ or at least NH$_2$ functional groups. The stretching modes of these transitions would appear close to the H$_2$O absorption (e.g. N-H stretching mode of NH$_3$ at 2.96~$\mu$m and amides in the region 2.70~--~3.30~$\mu$m). However, for NH$_3$ (at least), such transitions would also be accompanied by features associated with the bending and umbrella modes of NH$_3$ at 6.16 and 9.00~$\mu$m, respectively \citep{Bott10}; these wavelengths lie outside the NIR region probed by these AKARI observations, so no further analysis is possible here. Finally, Type 1 objects display the ``standard'' H$_2$O band profile with an extended red wing; the distinct shoulder seen in Type 2 is still likely to be present (and thus the additional carriers and molecular species remain) but is now subsumed by effects in the broad red wing of the OH stretching absorption band. The average H$_2$O ice column density on these lines of sight is greater than those of Types 3 and 2. However, as the fits show, this increase in column density alone cannot account for the changes in the red wing profile between Types 1, 2, and 3. Some change in the scattering properties of the ice must also occur, e.g. grain growth and mantle growth occur concurrently.

\begin{figure*}[htb]
\plottwo{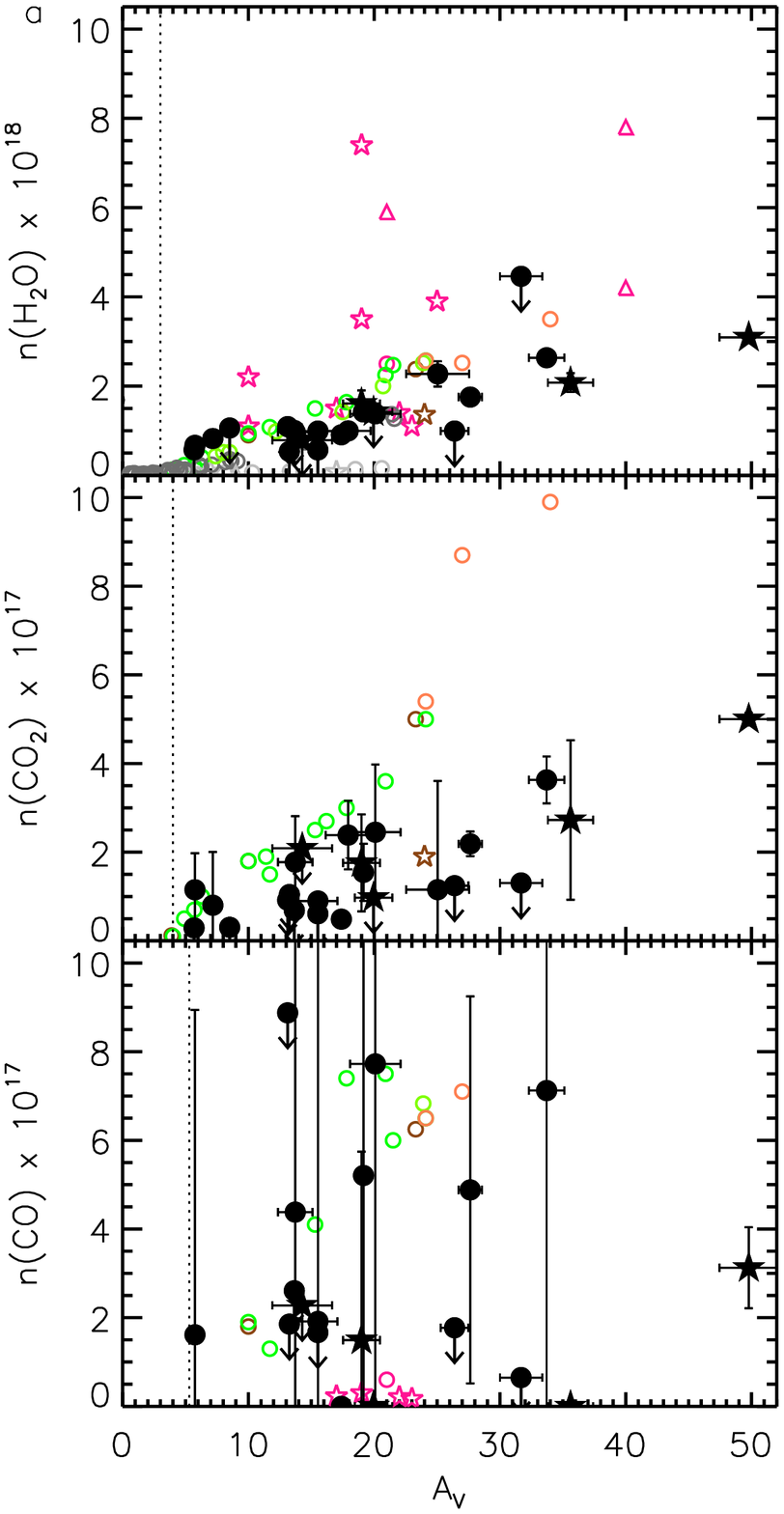}{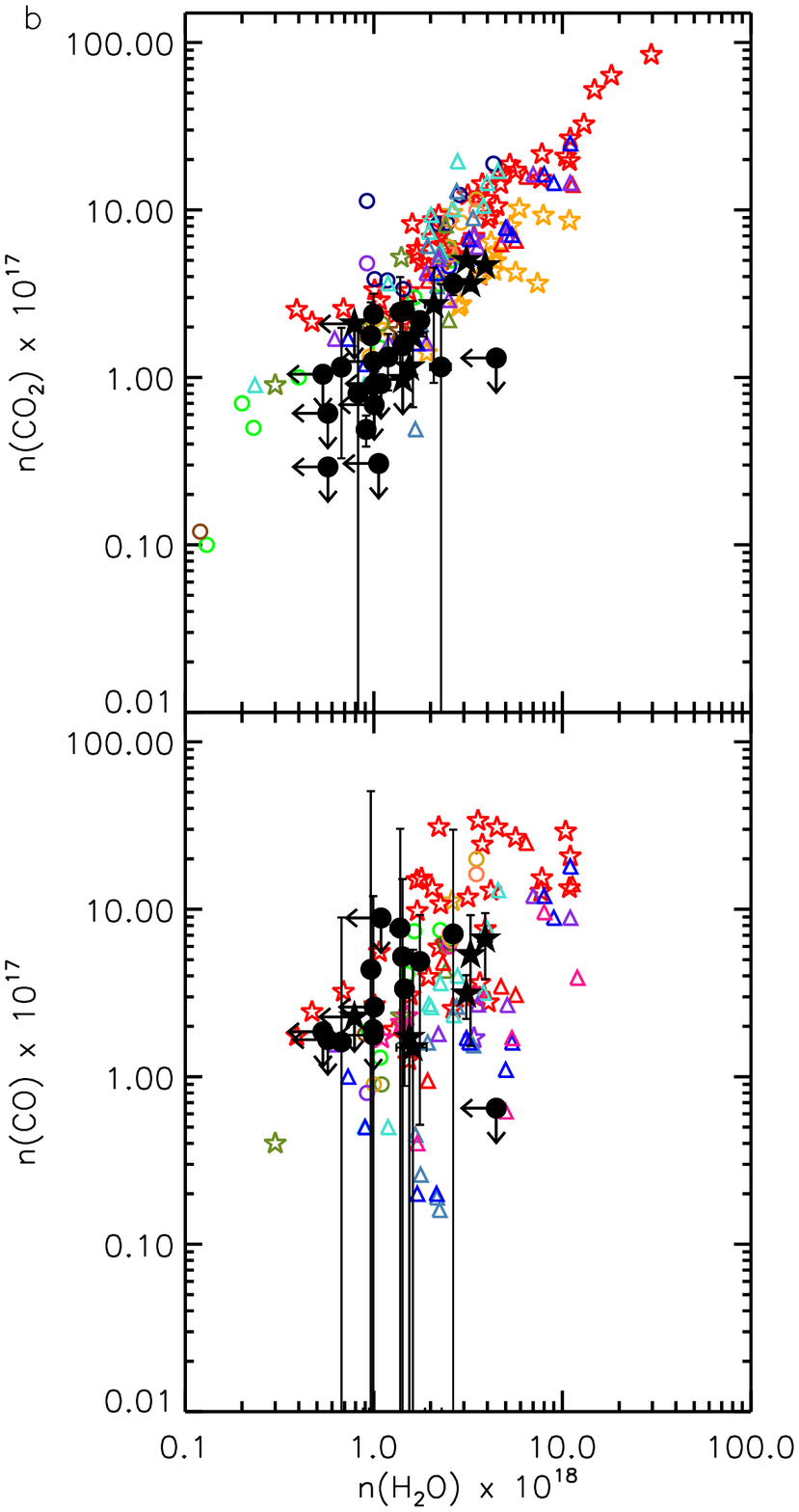} 
  \caption{Correlation plots of calculated H$_2$O, CO$_2$, and CO column densities with extinction. In this, and all subsequent, correlation plots, data calculated in this work are plotted as follows: background stars are plotted as filled circles ($\bullet$) and low mass YSOs as stars ($\star$), with upper limits designated by arrows ($\downarrow$). Also plotted, as empty symbols, are literature values for comparison. Intermediate and high mass YSOs from the literature have also been included for completeness, plotted as triangles ($\Delta$). In the electronic version, different studies are plotted in different colours, as defined below. Plot a) Column densities of H$_2$O, CO$_2$, and CO plotted against A$_V$. In each plot, the dotted line represents the critical A$_V$ value calculated by previous studies. Plot b) Column densities of CO$_2$ and CO plotted against H$_2$O. The literature values are taken from: light grey \citep{Whittet88}, royal blue \citep{ger99}, dark grey \citep{Murakawa00}, olive green \citep{num01}, violet \citep{gib04}, pink \citep{vanb04}, gold \citep{kne05}, brown \citep{ber05}, green \citep{Whittet07}, red \citep{pon08}, orange \citep{Zasowski09}, coral \citep{Whittet09}, turquoise \citep[extragalactic, ][]{shi10}, navy blue \citep{Boogert11}, blue steel \citep[extragalactic, ][]{oli11}, bright green \citep{Chiar11}.\label{fig-Av/H2O}}
\end{figure*}

The profile of the red wing on lines of sight towards YSOs is usually attributed to grain growth, accompanied by accelerated freeze-out, which is known to occur in dense cores and regions surrounding YSOs, where the density is high enough for the freeze-out timescale to become shorter than the age of the core \citep[e.g.][]{Kandori03,Roman07,Chapman09}.
It is very interesting, therefore, that Type 1 includes both YSOs and background stars, although the reasons for this have not yet been fully ascertained. One possibility is that background stars exhibiting this H$_2$O band profile probe lines of sight through regions of a core that are close to an existing YSO, and are thus probing ices and grains which are at a later evolutionary stage than those probed by background stars exhibiting Type 2 profiles. However, from our data this does not seem to be the case. Most of the background star subset of Type 1 are either isolated objects, or lie in fields of view containing only background stars. In the only fields of view containing a YSO and a background star (L~1165, Objects 28 (YSO) and 29, and LM226, Objects 26 (YSO) and 25), the pattern is unclear, as Object 29 is of Type 1, but Object 25 is of Type 2. Perhaps, rather, the background stars in Type 1 also probe regions which are more evolved, and are about to undergo star formation, suggesting that proximity to a YSO is not the only observational indicator of evolutionary state.

 A number of studies have tried to reproduce the extended red wing profile of the H$_2$O band using Mie scattering with an encapsulated grain, changing grain size and mantle thickness \citep{smith1989,Thi06,Dartois02}. Given that our Type 3 and 2 profiles do fit with a CDE grain shape model, we opted to continue using the CDE fitting with the Type 1 profile. Perhaps a scattering model which took into account the evolution of the ice mantle as well as grain growth would be able to more accurately model the changing profile of H$_2$O bands during the evolution from a molecular cloud to a YSO.

The most obvious and widely published candidate molecule to account for the H$_2$O band shoulders in Types 2 and 1 is CH$_3$OH. It has previously been observed towards multiple lines of sight including both low mass YSOs and background stars \citep{Brooke99,pon03b,Boogert11}. As the focus of this article is the analysis and quantification of the H$_2$O, CO$_2$ and CO ice absorption features, further analysis of the H$_2$O red wing was not performed. This issue will be addressed in a future article (Suutarinen et al, in preparation).

\subsection{CO}

The original aim of these observations with AKARI was to determine the relative column densities of H$_2$O, CO$_2$, and CO ices, with a view to mapping ices across molecular clouds. Given the low spectral resolution of AKARI, it was not anticipated that the absorption features would be fully resolved, and thus a component fitting method was not envisioned. However, the profiles of the CO bands observed are well enough resolved that fitting components is viable.

The fitting methodology for CO was developed by consideration of various possible methods, the most crucial of which are illustrated in Figure~\ref{fig-COgg1}. The primary CO feature (2160 -- 2120 cm$^{-1}$) contains only nine or ten data points, while the AKARI instrumental profile contains 10 data points. These data are thus below the Nyquist limit for data sampling, and we expect the CO feature to be undersampled.
A single component fit of pure CO to the primary CO feature using a lab spectrum (with applied CDE model, AKARI instrumental profile and resolution correction) (Figure~\ref{fig-COgg1}a) was discounted as the wings were poorly fitted. A two component fit of pure CO and CO in a water environment (middle component, CO$_{mc}$, and red component, CO$_{rc}$, \citet{pon03}), similar to that first described by  \citet{tie91}, well described the main CO feature, especially the wings (Figure~\ref{fig-COgg1}b), with the reduced chi-squared statistic improving for 74~\% of spectra. However, it did not account for the 2160~--~2180 cm$^{-1}$ feature, nor the slight excess in the red wing of the CO feature. Previous AKARI observations showed that gas phase CO rovibrational lines might also be present concurrently with the CO ice feature \citep{Aikawa12} at least on lines of sight towards the edge-on disks they observed. At the AKARI resolution, these gas phase lines are not resolved, but the possibility of major gas phase CO features in absorption was considered, and rejected, based upon a series of model gas phase spectra, one of which is shown in Figure~\ref{fig-COgg1}c. Clearly the gas phase features peak where the ice features do not, even at the AKARI resolution, and they cannot account for the extended red wing on the AKARI CO ice features. We conclude that there is very little, if any, gas phase CO contribution in our spectra, and do not include a gas phase component in our fit. The feature between 2160~--~2180 cm$^{-1}$, which is usually attributed to OCN$^{-}$ \citep[e.g.][]{pon03}, is included in the overall fit for completeness, although the minor components contributing to this feature are not discussed further in this work. In our spectra, the feature can usually only be fully accounted for by including both an OCN$^-$ component and the so-called CO$_{gg}$, CO gas-grain, feature \citep{fraser05}, as illustrated by \citet{vanb05}. CO$_{gg}$ represents gas phase CO that chemisorbs directly to the bare grain surface during or before formation of the underlying H$_2$O ice mantle \citep{fraser05}. As can be seen from Figure~\ref{fig-COgg1}d, the OCN$^{-}$ feature alone is not able to fully complete the fit and so, as illustrated in Figure~\ref{fig-COgg1}e, we use both components here. The inclusion of the secondary feature (2160~--~2180 cm$^{-1}$) to the fit makes no difference to the calculated CO column densities for all spectra from which CO values were calculated (nor for 80~\% of upper limit values).

Ultimately, our CO ice band fitting was based on a modified version of the well-established four component analysis of \citet{pon03}. In our analysis, CO$_{rc}$ is replaced by a Gaussian calculated from laboratory data \citep{cot03}, but CO$_{mc}$ is retained as in \citet{pon03} (a CDE-corrected Lorentzian at 2139~cm$^{-1}$). The third CO component from \citet{pon03} is omitted, not because it is not necessarily present, but because, given the AKARI resolution, there is no need to add additional components to a fit, where two components suffice. That only two components are required to fit the main CO feature (CO$_{mc}$ and CO$_{rc}$) supports our assertions that ARF, and in particular the stacking method, was appropriate for analysis of these data.  The 2160~--~2180 cm$^{-1}$ feature,  (the fourth component in the \citet{pon03} method), was fitted by a Gaussian derived from laboratory OCN$^{-}$ spectra by \citet{vanb05}. In addition, a laboratory spectrum of CO gas chemisorbed on a zeolite surface (\citep[CO$_{gg}$, ][]{fraser05}) was added to the model, to fit the outlying absorption feature centred at 2175~cm$^{-1}$. CO$_{mc}$ is CDE corrected to account for grain shape effects, and the sum of all components is corrected for the AKARI instrumental line profile. One key element of the multi-component fitting in \citet{pon03} is that all the components are used to fit all the observed spectra, changing only the relative intensities of each component from one line of sight to another. This is the premise we also adopt here.

The fits of CO are presented in Figure~\ref{fig-od-3a}.  Where there is no data in a plot, it is because no data was extracted in this region of the spectrum (e.g. Object~1). Where there is only a partial data extraction, the plot is shown, but no fit was made to the data (e.g. Object~11). Interestingly, it is very clear from Figure~\ref{fig-od-3a} that in at least two objects -- Objects 4 and Object 20 -- the CO ice main peak is fully described by a single pure CO component (CO$_{mc}$). In fact as Table~\ref{tbl-3} shows, up to nine objects fit into this category. Conversely, in Objects 8 and 18, the CO ice is fully fitted by only the CO$_{rc}$. However both these objects provide only upper limits on the CO ice column densities, and the identification of a single CO ice component is much less compelling than for Objects 4 and 20. Given that CO ice is formed in low temperature cores, where CO gas critically freezes-out as densities increase, it is perhaps not surprising to observe objects with only pure CO ice present. As the temperature of a core increases (usually due to the formation of a YSO), this pure CO ice is expected to either desorb or migrate into the pores of the underlying water ice, giving rise to the CO$_{rc}$ \citep{collings03}. In the laboratory it is only possible to detect the spectral feature associated with the CO$_{rc}$ at elevated temperatures \citep{collings03b}, but it seems unlikely that the lines of sight probed here are particularly warm. The origins of the CO$_{rc}$ therefore remain debatable, as illustrated recently by \citet{Cuppen11}.

\begin{figure}
\plotone{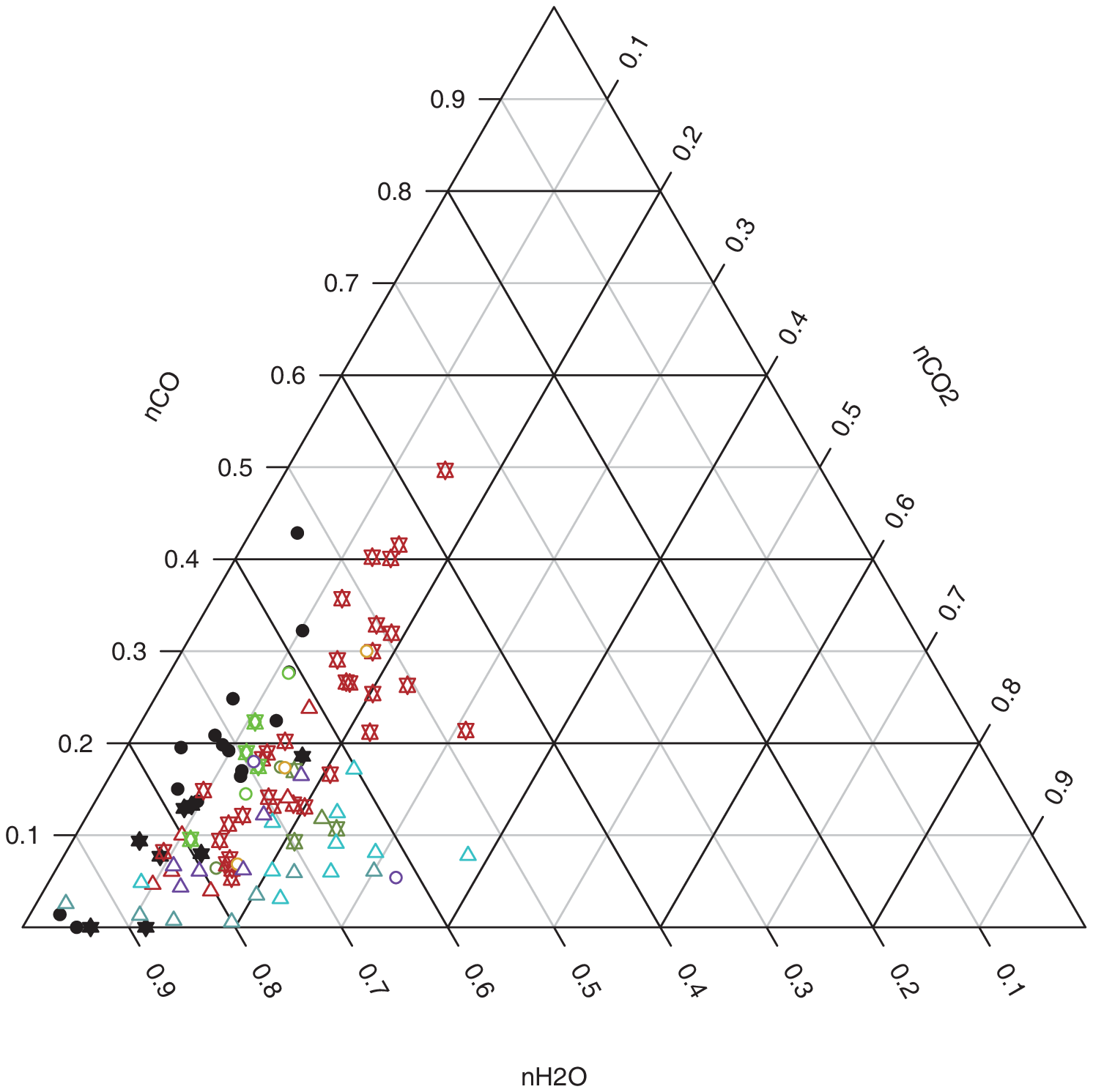}
  \caption{Ternary plot of the relative calculated column densities of H$_2$O, CO$_2$, and CO. For each line of sight, the sum of all three axes is 1. As before, in the electronic version the data labels are coloured as follows: olive green \citep{num01}, violet \citep{gib04}, gold \citep{kne05}, green \citep{Whittet07}, red \citep{pon08}, turquoise \citep[extragalactic, ][]{shi10}, blue steel \citep[extragalactic, ][]{oli11}. In this plot, low mass YSOs are plotted as sextiles (\ding{86}).\label{fig-ternary}}
\end{figure}

The column density of CO for the pure component, CO$_{mc}$, is calculated by:
\begin{equation}\label{eqn:co_mc}
N(CO_{mc}) = 6.03~cm^{-1} \times \tau_{max,mc} \times A^{-1}_{bulk},
\end{equation}
where A$_{bulk}$ = 1.1~$\times$~10$^{-17}$~cm molec$^{-1}$ is the band strength of the bulk material, and $\tau_{max,mc}$ is the optical depth at maximum absorption (Equation~3, \citet{pon03}). N(CO$_{rc}$) and N(OCN$^-$) are calculated using Equation~(\ref{eqn:colden}), with band strengths A$_{rc}$ = 1.1~$\times$~10$^{-17}$~cm molec$^{-1}$ \citep{gerakines95} and A$_{OCN^-}$ = 1.3~$\times$~10$^{-16}$~cm molec$^{-1}$ \citep{vanb04}. The column density of the CO gas-grain population is calculated by:
\begin{equation}\label{eqn:co_gg}
N(CO_{gg}) = 1.07~cm^{-1} \times \tau_{max,gg} \times A^{-1}_{bulk},
\end{equation}
where A$_{bulk}$ = 4.0~$\times$~10$^{-19}$~cm molec$^{-1}$ is the band strength of the bulk material \citep{fraser05}, and $\tau_{max,gg}$ is the optical depth at maximum absorption. All column densities calculated for CO$_{rc}$, CO$_{mc}$, OCN$^-$, and CO$_{gg}$ are presented in Table~\ref{tbl-3}. It should be noted that A$_{bulk}$ is calculated based upon laboratory spectra of a particular ice, using estimations for the parameters of ice thickness and number of molecules on the surface. If astrophysical ices differ in mixture or concentration, the relationship between optical depth and molecular abundance will change, and thus the relative concentration of one component to another will differ with respect to the values calculated here. This caveat holds for any interstellar ice column density analysis.

OCN$^-$ was believed to be formed around YSOs, based upon upper limits established by studies of background stars \citep{Whittet01} which suggested that OCN$^-$ is not found in quiescent clouds. However, more recent studies have revealed stricter upper limits on OCN$^-$ towards background stars probing quiescent regions \citep{kne05}. As is evident with reference to Table~\ref{tbl-3}, an OCN$^-$ component was fitted towards very few lines of sight (five YSOs and two background stars, all but one of which were upper limits). 

\subsection{CO$_2$}\label{sec-co2}

The CO$_2$ stretching mode is a very deep absorption feature and therefore it can easily become saturated; it is suspected that this is the case here. A second challenge in fitting the CO$_2$ ice region at the AKARI resolution limits, is that across the whole feature there are only 18~--~19 data points, which means that the line is not necessarily fully sampled, and potentially that the observed peak position in the spectrum is not coincident with the actual peak absorption. This also means that it is impossible to fit the CO$_2$ bands with the wings only, unlike in the case of H$_2$O. Furthermore, like the CO$_2$ bending mode at 15.2~$\mu$m, the stretching mode of CO$_2$ is very sensitive to the ice mixture, temperature, and other properties, as demonstrated by a number of laboratory experiments \citep[e.g.][]{vanb06}. Consequently, when fitting the CO$_2$ stretching mode there are a wide range of potential fitting parameters and degenerate solutions.

It is clear from previous AKARI studies that quantification of CO$_2$ abundances using AKARI spectra is challenging. \citet{shi10} calculated the CO$_2$ abundance towards massive YSOs in the Large Magellanic Cloud as $\sim$~36~\% relative to H$_2$O using a curve of growth method. In observations of CO$_2$ ice on lines of sight towards edge-on disks, \citet{Aikawa12}, determining abundances from the integration of a fitted laboratory spectrum, found CO$_2$/H$_2$O ratios broadly consistent with those published previously \citep[e.g.][]{pon08}. Of course, such differences may be wholly attributable to the types of objects observed -- extragalactic YSO sources in the case of \citet{shi10}; ice in edge-on YSO envelopes and disks in the case of \citet{Aikawa12}, where the collapse may have occurred so rapidly that less CO freeze-out led to less CO$_2$ formation. In the two cases cited, CO$_2$ column densities were calculated using a single laboratory component or a curve of growth method. In this work, we will fit our data with a two component model, to make better fits to the CO$_2$ stretching absorption.

\begin{figure*}[htb]
\plotone{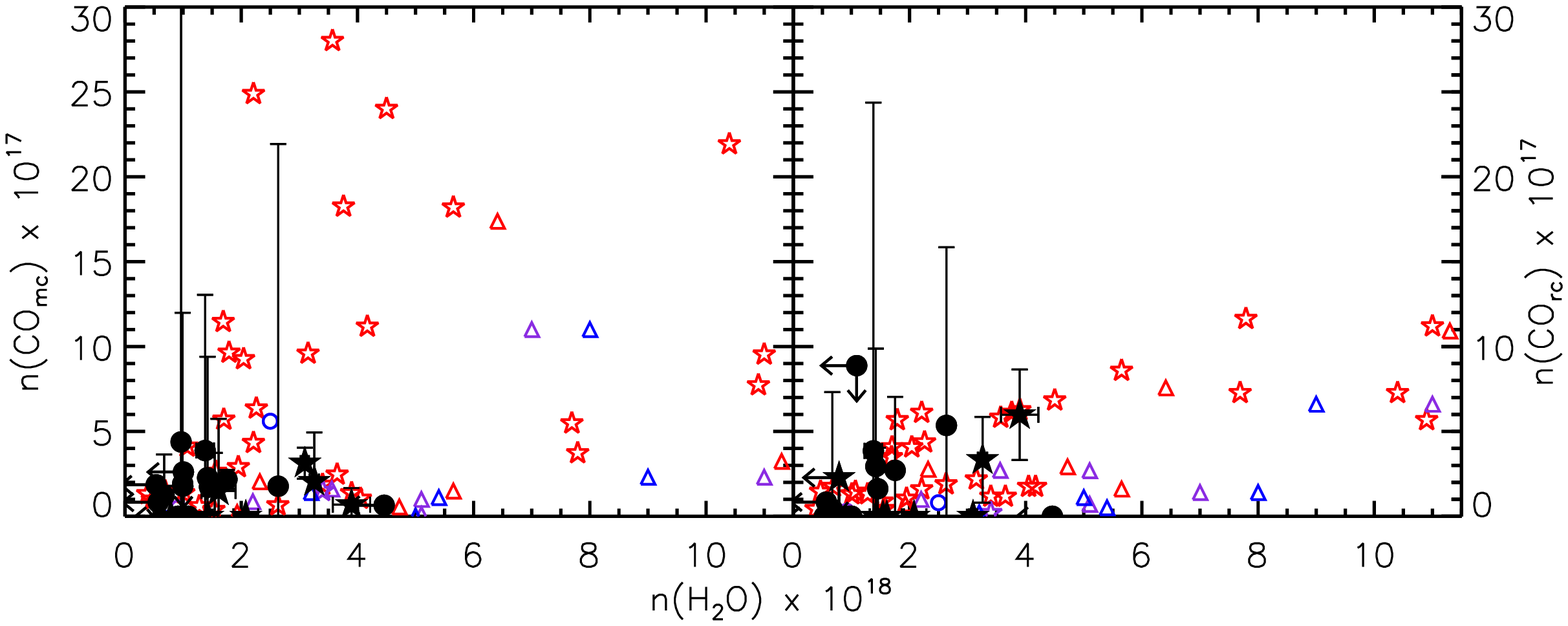} 
  \caption{Correlation plots of CO components with H$_2$O. \citep{ger99,gib04,pon08}. \label{fig-COcomponent_vsH2O}}
\end{figure*}

Given that previous detailed \emph{Spitzer} observations have established, from the bending mode feature, that CO$_2$ exists in two environments (a water-rich ice and a carbon monoxide-rich ice) towards YSOs \citep{pon08} and background stars \citep{kne05,Boogert11}, it was decided to fit the feature with two laboratory spectra: a CO$_2$:H$_2$O 14:100 ice at 10~K \citep{gerakines95} and a CO$_2$:CO 1:1 ice at 15~K \citep{Fraser04}. A Kramers-Kronig analysis \citep{boh83,ehr97}, followed by a CDE grain shape model and the AKARI instrumental line profile were applied to both laboratory spectra before fitting the peak (excluding the central four data points, as they might not fully describe the maximum absorption). From Figure~\ref{fig-od-2a}, it is clear that the strategy fits the observed data well, including the peak. Where there is saturation of the CO$_2$ band, this is also reflected by the fitting, with the wings well described but the final fit illustrating the significant undersampling in the AKARI data. The reduced chi-squared statistics were improved for all but one spectrum (97~\%) when comparing a two-component fit to fitting a single laboratory spectrum, further justifying our choice of fitting methodology.

Column densities were calculated using Equation~(\ref{eqn:colden}) with A~=~7.6~$\times$~10$^{-17}$~cm\,molecule$^{-1}$ \citep{gerakines95}, and are presented in Table~\ref{tbl-3}. For the YSOs in our sample the values obtained are commensurate with those previously reported for low-mass YSOs \citep{pon08,Aikawa12}; the column densities for the lines of sight towards background sources are generally lower, as found in previous studies \citep[e.g.][]{kne05,Boogert11}.

\section{Astrochemical implications}\label{section-correlations}

As explained in the introduction to this article, a key aim of these observations was to explore the links between H$_2$O, CO$_2$, and CO ice column densities, and to test the chemistry relating these species. Figure~\ref{fig-Av/H2O}a shows the calculated column densities of the three most abundant solid phase molecular species, H$_2$O, CO$_2$, and CO, plotted against visual extinction, A$_V$. The lowest A$_V$ of any object in these AKARI data is 5.68. 

Our results clearly follow the trends established by previous observations. Both the H$_2$O and CO$_2$ column densities increase with A$_V$, commensurate with the previously calculated critical A$_V$ values of $\sim$~3 and $\sim$~4, respectively, although such low extinctions were not probed in this subset of AKARI observations. For both H$_2$O and CO$_2$ we would expect an increase in ice column density with A$_V$; more dust equals, on average, more ice.

Conversely, in Figure~\ref{fig-Av/H2O}a, there is no clear relationship between CO and A$_V$. There is a much greater spread in CO values compared to those seen for H$_2$O and CO$_2$ in the panels above. CO is known to form in the gas phase and freeze-out onto the grain surface above a critical A$_V$ (which has been estimated, variously, as $\sim$~3 in Serpens \citep{Chiar94}, $\sim$~5 in Taurus \citep{Whittet89}, and $\sim$~11 in $\rho$~Ophiuchi \citep{Shuping00}), but not necessarily as a function of dust density, as the CO gas density is more critical \citep{pon03}.  Towards one line of sight, CO is observed in the solid phase at relatively high column density for a low extinction (Object 10, A$_V$ = 5.75, N(CO) = 1.6~$\times$~10$^{17}$ molecules cm$^{-2}$), which may suggest that, with sufficient sensitivity, ices can be detected at lower A$_V$ values than previously reported in the literature. 

\begin{figure*}[htb]
\plotone{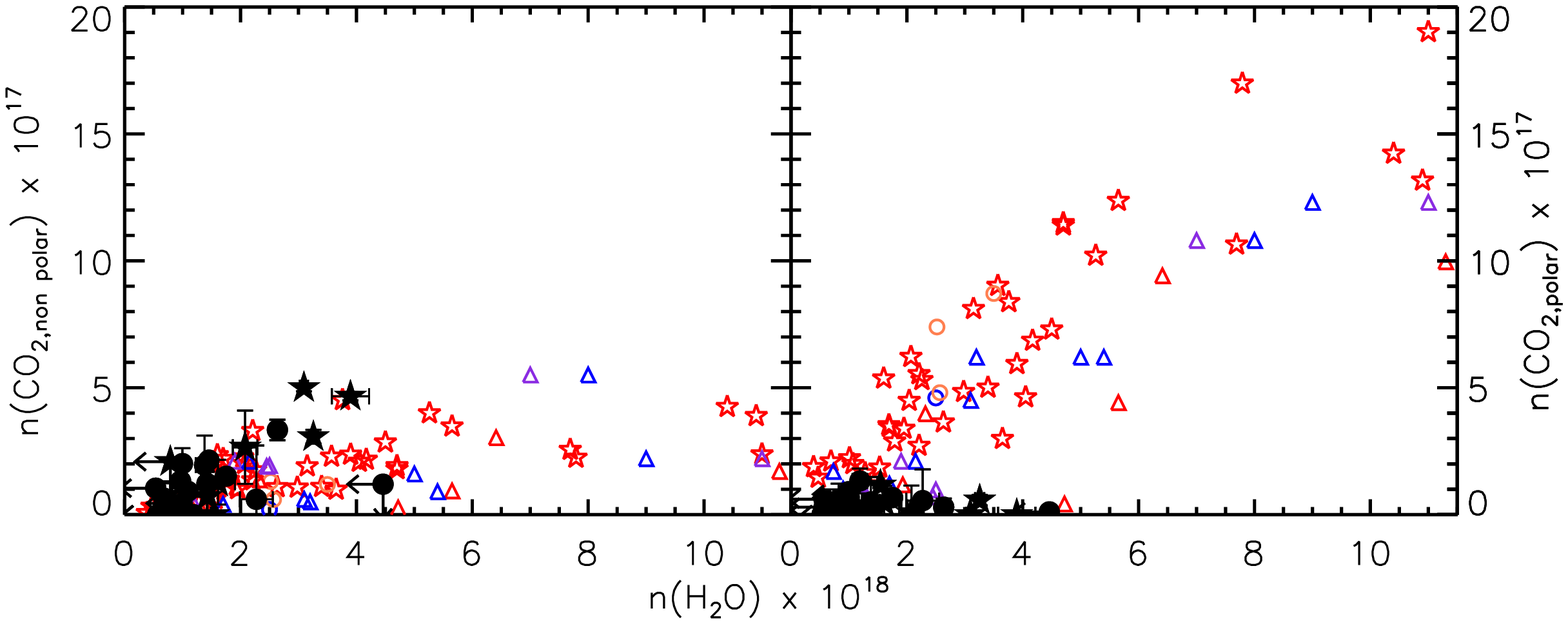} 
  \caption{Correlation plots of CO$_2$ components with H$_2$O \citep{ger99,gib04,pon08,Whittet09}.\label{fig-CO2component}}
\end{figure*}

Figure~\ref{fig-Av/H2O}b shows the column densities of CO$_2$ and CO plotted against the column density of H$_2$O. Overall, the data calculated in this study agree well with literature values; the relationships between H$_2$O, CO$_2$ and CO in molecular clouds and around YSOs are well established and it is important that new data continue to reinforce previous results. As expected, there is no clear relation between the H$_2$O and CO column densities, as the freeze-out and formation mechanisms of each species differ vastly. However, a linear correlation between the H$_2$O and CO$_2$ column densities is evident, indicating a potential link in the formation chemistry of these two species. It is possible that they both share a common chemical starting point; one route to H$_2$O formation involves H + OH \citep[e.g.][]{Cuppen07}, one route to CO$_2$ formation involves CO + OH \citep{Oba10,iop11a,nob11,zins11} and could account for the CO$_2$ ice observed in quiescent regions \citep{num01,Pontoppidan06}. Recent experimental results show that the relative rates of these two reactions, if occurring concurrently, and if they were the only reactions producing H$_2$O and CO$_2$, would generate around 20 times more H$_2$O than CO$_2$, assuming that all the reagents were equally available \citep{nob11}. Obviously, these conditions may not be satisfied even in a quiescent line of sight. Figure~\ref{fig-Av/H2O}b shows that the H$_2$O:CO$_2$ ratio is approximately 10:1, suggesting that other mechanisms must produce the remainder of the observed CO$_2$.  Unlike the subset of background stars and low luminosity YSOs surveyed here, column density values for low and high mass YSOs tend to be much higher \citep[e.g.][]{gib04,pon08}; corroborating evidence that UV processing or longer timescales promote CO$_2$ formation via more complex CO$_2$ formation mechanisms.

Figure~\ref{fig-ternary} is a ternary plot showing the relative column densities of all three species on a single plot. It includes fewer data points than Figure~\ref{fig-Av/H2O}, simply because far fewer studies have been able to ascertain the abundances of all three species towards individual lines of sight. This figure clearly illustrates the range of background star data and column densities which were probed by this survey.

Looking more closely at the three major ice species, more can be learnt about their interlinking chemistries by considering the individual components of each. During this process, it is important to consider all available literature data in order to ascertain general correlations, rather than potentially drawing conclusions based on survey-specific trends. We first consider the relationship between H$_2$O column density and the column densities of pure CO (CO$_{mc}$) and CO in a H$_2$O-rich ice environment (CO$_{rc}$) in Figure~\ref{fig-COcomponent_vsH2O}.

\begin{figure}[htb]
\plotone{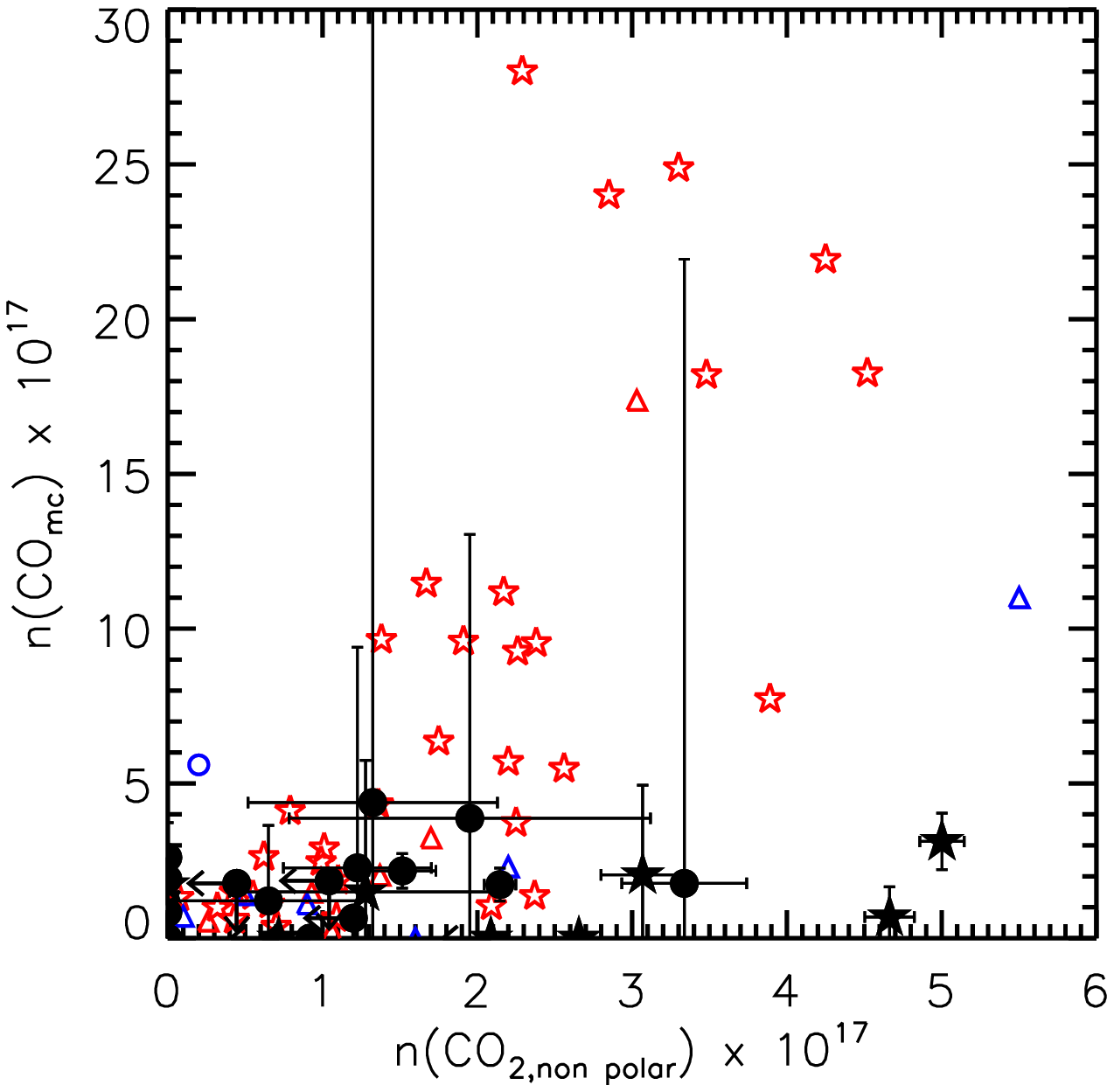} 
\caption{Correlation plot of CO$_{mc}$ column density plotted against CO$_2$ in a CO-rich ice environment \citep{ger99,gib04,pon08}.\label{fig-COvsCO2}}
\end{figure}

It is postulated from laboratory experiments that these two CO ice environments promote very different ice chemistries: in the pure CO ice the reactions are dominated by successive H atom, O atom and radical additions to the CO molecule to form CO$_2$, HCOOH, H$_2$CO and CH$_3$OH \citep[e.g.][]{wat04,Fuchs09}; in the water-rich CO ice, photon- and electron-induced chemical processes become more important, firstly because the CO molecule is trapped, leading to a CO chemistry at temperatures above its desorption energy, and secondly because (unlike the CO molecule) H$_2$O can be photodissociated, allowing reaction with CO to form CO$_2$, HCOOH, H$_2$CO and CH$_3$OH as well as many more complex molecules \citep[e.g.][]{ger96,Jamieson06}. It is only through observational evidence that we can start to discriminate between these processes and refine gas-grain astrochemical models. 

It is clear from analysis of Figure~\ref{fig-COcomponent_vsH2O} that there is no correlation of H$_2$O with pure CO, which condenses out of the gas phase as a layer on top of the H$_2$O. However, for the H$_2$O-rich phase of the CO, as expected, N(CO$_{rc}$) increases with N(H$_2$O). \citet{collings03} offer evidence for the porosity of H$_2$O in the ISM, as their experiments suggest that the population of CO in a H$_2$O-rich environment develops upon heating of a layered system. A higher column density of H$_2$O ice suggests a larger number of pores for the CO to migrate into with time or temperature to produce a mixed phase \citep{pon03}. As Figure~\ref{fig-COcomponent_vsH2O} shows, towards YSOs there is generally a higher column density of CO in a H$_2$O-rich environment. The ice has been present in these regions for longer, so it is therefore more likely that the CO has had the time (or the temperature change) necessary to migrate into a mixed H$_2$O layer. Otherwise, background stars would be expected to probe lines of sight with equally high column densities of CO$_{rc}$. It is important to note that CO$_{rc}$ is a derived feature, as it can not be measured in isolation in the laboratory.

\begin{figure*}[htb]
\plotone{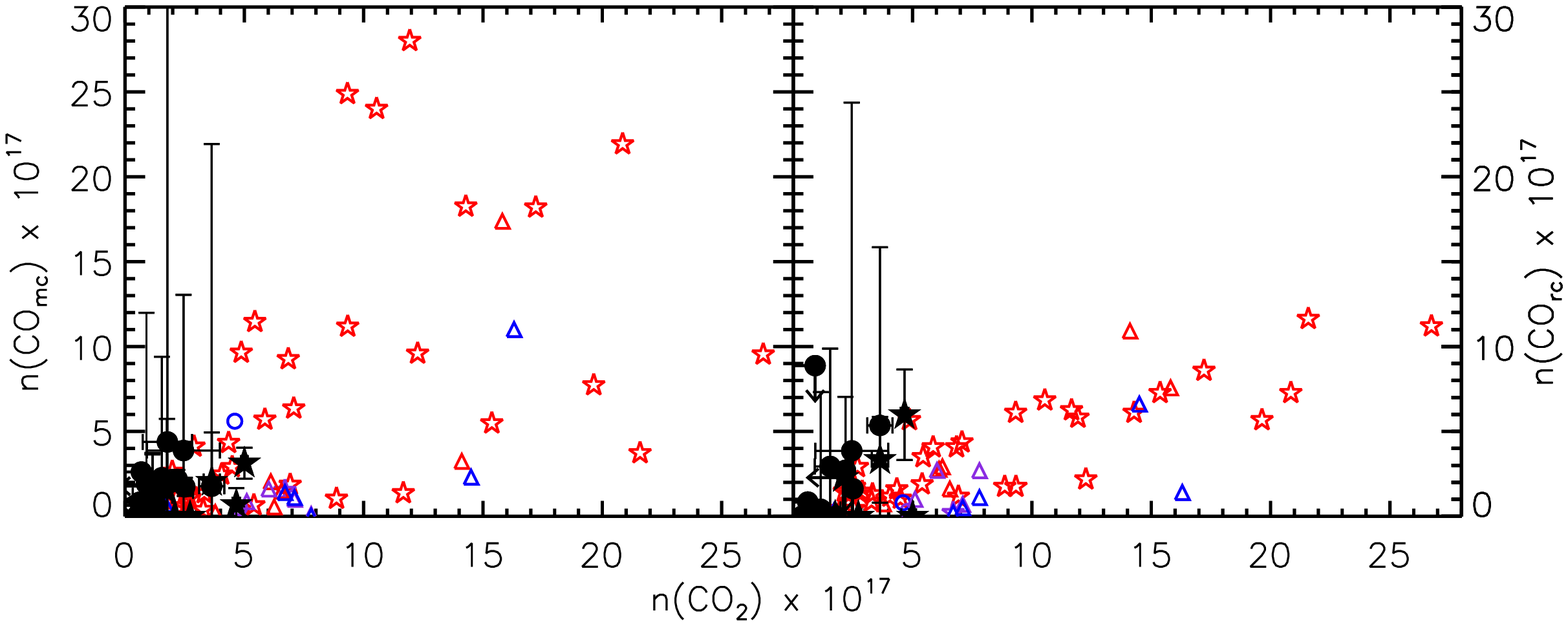}
  \caption{Correlation plots of CO components with CO$_{2, polar}$ \citep{ger99,gib04,pon08}.\label{fig-COcomponent_vsCO2}}
\end{figure*}

Figure~\ref{fig-CO2component} illustrates the relationship between the H$_2$O column density and that of the two components of CO$_2$ -- CO$_2$ in a CO-rich ice (CO$_{2, non\, polar}$) and CO$_2$ in a H$_2$O-rich ice (CO$_{2, polar}$). When the component column densities from the broader literature, primarily from low and intermediate mass objects, are included in the plot (open symbols), two different correlation gradients are clearly evident, which both scale directly with ice abundance (albeit at different rates) suggesting that the dominant formation mechanisms for CO$_2$ in  water-rich and water-poor environments are independent of one another. \citet{Whittet09} generated similar gradient profiles by comparing the total elemental O atom distribution, derived by summing the CO, CO$_2$, and H$_2$O ice column densities, in the H$_2$O-rich (\emph{polar}) and CO-rich (\emph{apolar}) ice phases on extincted lines of sight towards a sample of around ten bright background stars and YSOs. They show that the elemental O is uniformly distributed between the water-rich and water-poor ice components, and is well correlated with the dust density, suggesting that the final fraction of O atoms accumulating in interstellar ices is a constant reservoir from cloud to cloud, but that the absolute source to source O atom distribution between the three major ice molecules -- CO, H$_2$O, and CO$_2$ -- can vary significantly. They attribute most of this variance to the CO$_2$ population in the water-rich ice component, suggesting that the CO$_2$ ice column densities are invariant with the efficiency of surface chemical pathways, instead reflecting variations in gas phase chemistry (related to CO gas densities) or photodesorption of the CO ice.

Our data do not entirely corroborate this conclusion. The component column densities of CO$_2$ in a CO-rich ice (CO$_{2, non\, polar}$), derived from our two-component fitting, match very well with previously reported values (as is the case for the CO ice components in Figure~\ref{fig-COcomponent_vsH2O}), providing reassurance in our analysis methods, despite the fitting limitations imposed by the AKARI resolution discussed in \S~\ref{sec-co2}. It is therefore interesting to note that we have sampled a very specific population of CO$_2$ in a H$_2$O-rich ice (CO$_{2, polar}$), which clearly occupy the lower left-hand region of Figure~\ref{fig-CO2component}, commensurate with a handful of previous data points. It seems therefore that the formation of CO$_2$ in H$_2$O-rich ices is actually bimodal. The most likely scenario is that CO$_2$ ices first form via atom and radical surface reactions of CO in competition with H$_2$O formation, producing a limited concentration of CO$_2$ in a H$_2$O-rich ice (CO$_{2, polar}$). Later, a combination of CO freeze-out and CO migration to the water-rich ice layer, followed by energetic processing of the ices, leads to a second phase of CO$_2$ formation, which then dominates the CO$_2$ in a H$_2$O-rich ice (CO$_{2, polar}$) column density. This is entirely in agreement with the ``early'' and ``late'' formation mechanism model proposed by \citet{Oberg11}, and implies that the CO$_2$:H$_2$O ratio will vary significantly from source to source (as indicated by \citet{Whittet09}), but that this ratio will initially be dependent on competing surface reaction rates between CO, H, and OH to form CO$_2$ and H$_2$O, then subsequently on the degree to which ice mantles are thermally or energetically processed, rather than the prevailing gas phase conditions. \citet{poteet2013} have very recently illustrated that crystalline CO$_2$-rich ice layers can dominate grain mantles under the unusual circumstances where significant thermal processing occurs in the vicinity of a protostar (in this case HOPS-68), On the one hand, therefore, Figure~\ref{fig-CO2component} corroborates the \citet{Whittet09} hypothesis: the component of CO$_2$ ice in the H$_2$O-rich layer dominates the correlations between the major interstellar ice components. However, whilst our overall CO$_2$ column densities match well with existing literature values, the fraction of CO$_2$ ice we observe in water-rich environments is relatively small.

Previous studies have illustrated that there is not a direct link between CO freeze-out and all CO$_2$ formation, as there is no overall correlation in the abundances of total CO and total CO$_2$ \citep[e.g.][]{ger99}. In Figure~\ref{fig-COvsCO2}, the plot of the pure component of CO against the CO-rich component of CO$_2$ shows no correlation, suggesting that the formation route to CO$_2$ formed in the CO layer is not a simple one, and probably not driven only by CO reacting with O, H, or OH from the gas phase; a similar conclusion was reached from our analysis of Figure~\ref{fig-Av/H2O}b. It is likely that the relative CO and CO$_2$ populations in the CO-rich ice environments are additionally regulated by competing photo-desorption processes \citep[e.g.][]{oberg2007,oberg2009,fayolle2011}.

Figure~\ref{fig-COcomponent_vsCO2} shows plots of the column density of the water-rich CO$_2$ component versus the column density of the CO components CO$_{mc}$ and CO$_{rc}$. It is very clear from these plots that there is a strong link between the column densities of CO$_2$ and CO in a H$_2$O-rich environment (CO$_{rc}$). Thus, some CO$_2$ must form from CO in a H$_2$O environment. Mirroring the relationship established between H$_2$O and CO$_2$ in a H$_2$O-rich environment (Figure~\ref{fig-CO2component}), background stars probing quiescent lines of sight have the lowest column densities of CO in a H$_2$O-rich environment, while in the more evolved YSOs, there is a higher column density, suggesting that CO$_2$ forms from CO which has migrated into the H$_2$O upon formation of a star in the core. This is key evidence supporting the conclusions of the ice mapping in \citet{Pontoppidan06}.

These data illustrate the links between all of the components present in the CO and CO$_2$ absorption features, and the sequence of CO$_2$ formation in molecular clouds and star-forming cores. The proposed sequence is outlined here: initially, CO$_2$ forms concurrently with H$_2$O, producing an abundance of CO$_2$ in H$_2$O. Upon critical freeze-out of CO, further CO$_2$ is formed by reactions of CO with O, H, and OH, producing a CO$_2$ component in a CO-rich ice. With time and, potentially, processing, some CO migrates into the H$_2$O ice, where it can react to form large column densities of CO$_2$ in a H$_2$O-rich ice, likely by an energetic route enhanced by UV radiation from a newly formed YSO.

\section{Conclusions}

In this paper, the 2.5~--~5~$\mu$m spectra of 22 background stars and eight YSOs have been analysed to determine the column densities of key solid phase molecular species towards those lines of sight. Prior to this study, only one complete spectrum of a background star had been obtained in this wavelength region, a region containing absorption bands for the three most abundant molecules present in interstellar ice. Thus, the addition of 22 spectra towards background stars to the literature provides valuable data to the community, which could help to benchmark the initial conditions in star-forming regions before the onset of star formation. 

The analysis of this large sample of data by a rigorous fitting approach reinforces the argument that the methods employed here are possible for large datasets \citep[as previously shown by e.g.][]{pon03,pon08,Boogert08,Oberg11}. It is not necessary to employ a ``mix and match'' approach using multiple laboratory spectra in order to define the ice characteristics of a wide range of objects. 

In summary, the key findings of this survey are:

\begin{itemize}
\item The sensitivity of AKARI enabled us to probe lines of sight towards objects with fluxes $<$~5~mJy, allowing observation of spectra in a previously inaccessible parameter space.
\item CO$_2$ and H$_2$O were found to be ubiquitous where ice was observed, at column densities commensurate with previously reported literature values.
\item From the profile of the H$_2$O ice band, the extended red wing seems to be an evolutionary indicator of both dust properties and the extent to which the ice mantle covers the grain. 
\item New observations of the CO$_2$ stretching absorption band profiles, and ice column densities for lines of sight towards 21 background stars, have been obtained.
\item CO ice absorption features were extracted from the spectra of 23 of the 30 lines of sight probed. In at least two objects -- Objects 4 and 20 -- the CO ice absorption band profile is fully described by a single pure CO component only.
\item On 15 lines of sight towards background stars, the column densities of all three species H$_2$O, CO$_2$, and CO were extracted.
\item By considering the column densities of CO and CO$_2$ in different ice components, it is possible to postulate a CO$_2$ formation scenario involving two or three distinct stages. The proposed timeline is: early formation concurrent with H$_2$O; late formation by energetic routes when CO has frozen out and migrated into an H$_2$O environment; and the formation of CO$_2$ in the CO layer at an uncertain time, and by uncertain mechanisms.
\end{itemize}

As our knowledge of interstellar solid state ices increases, many common factors are continuing to emerge in our understanding of their formation and evolution. By surveying ices features in the spectra of faint YSOs and background sources, we have sampled a further region of parameter space in which ices can be detected. Clearly, even where correlations are evident between one ice species and another, the range of molecular column densities remains vast. The local environment must, therefore, have a significant effect on ice column density, composition and growth. Even the individual components of a species potentially vary across a cloud. In order to use solid phase molecular species as probes of ice and gas chemistry, it will be necessary to map their distributions on a 2D spatial scale, as well as making many more observations of both solid phase species and gas phase molecules.

\acknowledgments
The authors are very grateful to Y. Ohyama for fruitful discussions, and thank M. Tamura, M. Ueno, R. Kandori, and A. Kawamura for their help with the initial target selection of the ISICE programme.  J.A.N. is a Royal Commission for the Exhibition of 1851 Research Fellow, and acknowledges the financial support of the University of Strathclyde, the Scottish Universities Physics Alliance, and the Japan Society for the Promotion of Science. The research leading to these results has received funding from the European Community's Seventh Framework Programme FP7/2007-2013 under grant agreement No. 238258 (LASSIE). H.J.F. and J.A.N. are grateful to EU funded COST Action CM0805 ``The Chemical Cosmos: Understanding Chemistry in Astronomical Environments'' for a funding contribution towards the final production of this work.

{\it Facilities:} \facility{AKARI}.


\end{document}